\documentclass[preprint]{aastex}

\begin{document}
\title{Weak Turbulence in the HD 163296 Protoplanetary Disk Revealed by ALMA CO Observations}
\author{
Kevin M. Flaherty\altaffilmark{1},
A. Meredith Hughes\altaffilmark{1},
Katherine A. Rosenfeld\altaffilmark{2},
Sean M. Andrews\altaffilmark{2},
Eugene Chiang\altaffilmark{3,4},
Jacob B. Simon\altaffilmark{5,6,7},
Skylar Kerzner\altaffilmark{3},
David J. Wilner\altaffilmark{2}}
\altaffiltext{1}{Van Vleck Observatory, Astronomy Department, Wesleyan University, 96 Foss Hill Drive, Middletown, CT 06459}
\altaffiltext{2}{Harvard-Smithsonian Center for Astrophysics, 60 Garden Street, Cambridge, MA 02138}
\altaffiltext{3}{Department of Earth and Planetary Science, 307 McCone Hall, University of California, Berkeley, CA 94720}
\altaffiltext{4}{Department of Astronomy, 501 Campbell Hall, University of California, Berkeley, CA 94720}
\altaffiltext{5}{Department of Space Studies, Southwest Research Institute, Boulder, CO 80302}
\altaffiltext{6}{Sagan Fellow}
\altaffiltext{7}{JILA, University of Colorado and NIST, 440 HCB, Bouler, CO 80309}

\begin{abstract}
Turbulence can transport angular momentum in protoplanetary disks and influence the growth and evolution of planets. With spatially and spectrally resolved molecular emission line measurements provided by (sub)millimeter interferometric observations, it is possible to directly measure non-thermal motions in the disk gas that can be attributed to this turbulence. We report a new constraint on the turbulence in the disk around HD 163296, a nearby young A star, determined from ALMA Science Verification observations of four CO emission lines (the CO(3-2), CO(2-1), $^{13}$CO(2-1), and C$^{18}$O(2-1) transitions). The different optical depths for these lines permit probes of non-thermal line-widths at a range of physical conditions (temperature and density) and depths into the disk interior. We derive stringent limits on the non-thermal motions in the upper layers of the outer disk such that any contribution to the line-widths from turbulence is $<$3\%\ of the local sound speed. These limits are approximately an order of magnitude lower than theoretical predictions for full-blown MHD turbulence driven by the magneto-rotational instability, potentially suggesting that this mechanism is less efficient in the outer (R$\gtrsim30$AU) disk than has been previously considered.

\end{abstract}

\section{Introduction}
Planets form within the disks of gas and dust that surround young stars and are subject to the dynamics within these systems. In particular,  magneto-hydrodynamic (MHD) turbulence generated by the magneto-rotational instability (MRI), which is the leading theoretical mechanism driving angular momentum transport and disk accretion \citep[and references therein]{bal98,arm11,tur14}, can have a large effect on the planet formation process. Dust grains naturally gravitationally settle toward the midplane of the disk, with large grains settling faster than small grains. Turbulent motions could counteract this process \citep[e.g.][]{dub95,joh05,car06,cie07} preventing grains from collecting in high densities at the midplane and thereby limiting the rate of planetesimal formation. Laboratory measurements find that grains can stick together if they collide at less than $\sim$1-30 m/s, depending on the composition of the grain \citep{pop00,blu00,blu08,gun11} and enhanced turbulence could easily create relative motions that exceed this threshold. Turbulence may also allow pebbles and rocks to collect into overdense regions where they eventually collapse into planetesimals \citep{kla97,joh07,you02}. For planetesimals, turbulence may slow inward migration \citep{lau04} although at the expense of increased erosive collisions \citep{ida08}. The threshold for a massive planet to open a gap within the disk depends on the ability of the accretion flow to refill the gap; weaker turbulence, and hence a weaker accretion flow, would allow smaller planets to open large disk gaps \citep{cri06,kle12,fun14}. The variety of effects indicate that a detailed understanding of planet formation requires knowledge of the strength of the turbulent motion.

Much of the work investigating the strength of turbulence in disks has been strictly theoretical, focusing on the complex physics involved. Given the typical physical conditions within a protoplanetary disk, MRI is the predominant mechanism for angular momentum transport. The coupling of magnetic fields within the disk, along with rotational shear, drive the MRI that creates turbulence and leads to angular momentum exchange \citep{bal98}. In the ideal MHD limit instabilities grow quickly throughout the disk \citep{bal91}. Low ionization throughout much of the disk introduces various non-ideal MHD effects but independent of these complex effects numerical simulations find that MRI produces a strong vertical gradient in the strength of the turbulent motion \citep{mil00,flo11,sim15}. Close to the midplane the velocities are only a few percent of the local sound speed while at $z>$2-3H, where H is the pressure scale height, turbulent velocities approach the sound speed \citep{fro06,sim13,sim15}. Gravito-turbulence also predicts large velocities, although with a much different vertical gradient \citep{for12,shi14}.

Several groups have used observational methods to provide constraints on turbulence in order to inform the simulations. \citet{har98} find that viscously spreading disks with $\alpha\sim0.01$, where $\alpha$ relates the sound speed $c_s$ and pressure scale height to viscosity ($\nu=\alpha c_s H$), are consistent with the observed decrease in accretion rate with time and measurements of disk mass. This method does not directly measure turbulent motion, but instead relies on scaling relations among physically relevant parameters. More direct methods use the motion of the gas itself and  its influence on molecular emission lines, in particular the broadening of the line profile.  Line broadening consistent with transonic turbulent motion has been needed to fit near-infrared CO overtone emission, sensitive to the upper layers of the inner AU of the disk, in a number of sources \citep{car04,har04,naj09,naj96}. Sub-mm interferometric observations are also sensitive to line broadening, but can also trace other effects of turbulence such as changes in the peak-to-trough ratio of the spectral line profile, as well as broadening of the images \citep{sim15}. Resolved images also help break some of the degeneracies with e.g. Keplerian rotation that influence unresolved observations. \citet{hug11} model high spectral resolution Submillimeter Array (SMA) observations of CO(3-2) emission and find a tentative (3$\sigma$) detection of 300m s$^{-1}$ turbulence within the HD 163296 system but only an upper limit ($<$40m s$^{-1}$) in TW Hya. \citet{gui12} use CS, a heavier molecule that is less affected than CO by thermal broadening, to measure 130m s$^{-1}$ turbulence within DM Tau, which suggests turbulence of $\sim$50\%\ of the sound speed in the outer disk.

Of the three stars that have been studied in the submm, the HD 163296 disk hints at the strongest turbulent signal. HD 163296 (=MWC 275) is a nearby (122 pc), 2.3 M$_{\odot}$, $\sim3$ Myr old A1 star \citep{mon09} with a large circumstellar disk typical for its age. The size and proximity of this system make it a fruitful target for sub-mm studies exploring the chemistry and physical conditions of the cold outer disk \citep{qi11,til12,qi13a,qi13b,mat13}. Recently available science verification data from the Atacama Large Millimeter/submillimeter Array (ALMA) has opened up a new level of detail on the submm emission from the disk around HD 163296. These data spatially resolve a vertical temperature gradient within the disk \citep{ros13}. Detailed constraints on this structure help to mitigate one of the key degeneracies in earlier measurements of turbulence within this system \citep{hug11}. Here we use a physically sophisticated model, along with a Markov-Chain Monte-Carlo search method, to fit four emission lines (CO(3-2), CO(2-1), $^{13}$CO(2-1), and C$^{18}$O(2-1)) both together and separately to probe different layers within the disk. In sections~\ref{data},\ref{model} we describe the data, our model, and our fitting procedure. In sections~\ref{results},\ref{turbulence} we discuss the results of this fitting, where we find an upper limit on turbulence that is an order of magnitude lower than predicted by numerical simulations of MRI, as well as the implications regarding this weak turbulence. 

\section{The Data\label{data}}
ALMA observed the HD 163296 system as part of early Science Verification operations. Band 6 data were taken on 2012 June 9, 23, and July 7th when 24 12m antennas were available covering baselines of 20 to 400m. Juno, Neptune, and Mars were used as amplitude calibrators, while the quasar J1924-292 was used as the bandpass calibrator. Science observations were interleaved with measurements of the phase calibrator quasar J1733-130 with a total on-source integration time of 84 minutes. The spectral coverage was split into four spectral windows. Spectral windows 0 and 3 are composed of mostly line-free continuum while spectral window 1 included the $^{13}$CO and C$^{18}$O(2-1) lines and spectral window 2 included the CO(2-1) line. The four spectral windows were centered on 217.105, 219.969, 230.969 and 233.999 GHz with bandwidths of 1.875, 0.937, 0.937, and 1.875 GHz and spectral resolutions of 0.67, 0.33, 0.32, and 0.62 km s$^{-1}$. The beam size is 0.81x0.65'', while the primary beam has a FWHM of 22''.

Band 7 data were taken on 2012 June 9, 11, 22, and July 6 with the same antenna configuration as the band 6 data. Juno and Neptune were used as amplitude calibrators with the quasars J1733-130 and J1924-292 used as the phase and bandpass calibrators respectively. The total on-source integration time was 140 minutes. The four spectral windows were centered at 360.169, 356.734, 345.796, and 346.998 GHz with bandwidths of 117.3, 468.8, 468.8, 937.5 MHz and spectral resolutions of 0.025, 0.10, 0.10, 0.21 km s$^{-1}$. Spectral windows 0 and 3 cover line-free continuum while spectral window 1 contains HCO+(4-3) and spectral window 2 focuses on the CO(3-2) line. We do not include HCO+ in our analysis to simplify the chemical structure that must be accounted for in our model. The beam size is 0.67x0.43'' with a primary beam size of 14''. 

We use the calibrated Science Verification data; more details on the calibration procedure can be found on the NRAO website\footnote{http://almascience.nrao.edu/alma-data/science-verification}. We do not perform self-calibration which likely leads to an overestimate of the statistical uncertainties although, as discussed below, our results are dominated by systematic uncertainties and degneracies betwenn various model parameters rather than statistical uncertainties. For our analysis we use a low-resolution and high-resolution version of the CO(3-2) line. The low-resolution spectrum has been binned to a resolution of 0.3 km/s and includes only the central 15 channels. The high-resolution spectrum is at the native resolution of 0.1 km/s and included 101 channels covering the entire line profile. The J=2-1 lines were extracted from the calibrated data set, including only the channels with significant line emission (typically $\sim$30 channels).

While images are not used in model fitting, they are created to help interpret model fits. In creating images we clean the central 14x14'' with natural weighting and pixel sizes of 0.05''. When employed, the same cleaning routine is applied to both the model and the data to ensure consistent results.

We derive the uncertainties on the visibilities by calculating the the dispersion among the 70 visibility points with baselines closest to the particular uv point. This method accurately estimates the weight on each data point without detailed knowledge of the observing conditions, such as the system temperature. Seventy uv points serves as a balance between having enough information to accurately estimate the dispersion, without covering a large enough area in the uv plane so as to confuse real variations in the visibilities with noise. Based on the residuals between observed and modeled visibilities, normalized by the uncertainties, we find that this method accurately estimates the errors and that these errors are Gaussian, a necessary condition for using the chi-squared statistic when evaluating the likelihood of a model.

\section{The Model\label{model}}
To measure the turbulence within the disk, we employ a simple parameteric model to fit the data. We start with a temperature and surface density structure, compute the vertical structure using hydrostatic equilibrium, calculate the velocity field taking into account Keplerian rotation as well as pressure support, and compute the radiative transfer of the line through the disk, accounting for both thermal and turbulent broadening. 
This model is based on the work of \citet{ros12} and \citet{ros13}, itself based on the structure laid out in \citet{dar03}, and has successfully been used to model CO emission from other disks \citep{ros12,and12,ros13b,wil14,ros14}. We choose not to derive the gas temperature structure based on a radiative transfer calculation accounting for stellar irradiation and gas/dust interactions in order to speed up the model calculations and allow for a Markov-Chain Monte-Carlo (MCMC) approach for constraining the model parameters (described below).

\subsection{Temperature, Density and Velocity Structure}

In hydrostatic equilibrium, the gas density and temperature are related by:
\begin{equation}
-\frac{\partial \ln \rho_{\rm gas}}{\partial z} = \frac{\partial \ln T_{\rm gas}}{\partial z} + \frac{1}{c_s^2}\left[\frac{GM_*z}{(r^2+z^2)^{3/2}}\right],
\end{equation}
where $\rho_{\rm gas}$ is the gas density, $T_{\rm gas}$ is the gas temperature, $M_*$ is the stellar mass and $r,z$ are the radial and vertical position within the disk. The sound speed is given by $c_s^2=k_BT_{\rm gas}/\mu m_h$ where $k_B$ is Boltzmann's constant and $m_h$ is the mass of hydrogen. The disk is assumed to be made of 80\%\ molecular hydrogen with $\mu=2.37$. 

At each radius the vertical density structure is calculated from the temperature profile. For the temperature structure we use functional form laid out in \citet{ros13},
\begin{equation}
T_{\rm gas}(r,z) = \left\{
\begin{array}{ll}
T_{\rm atm} + (T_{\rm mid}-T_{\rm atm})(\cos\frac{\pi z}{2Z_q})^{2\delta} & \mbox{if $z < Z_q$} \\
T_{\rm atm} & \mbox{if $z \ge Z_q$}
\end{array},
\right.
\end{equation}
, which is based on the \citet{dar03} Type II structure and has been used to successfully model the HD 163296 disk CO emission previously. This temperature profile increases from $T_{\rm mid}$ at the midplane to $T_{\rm atm}$ at a height of $Z_q$. The midplane remains cold because it is shielded from the stellar flux that heats the surface layers. Mechanical heating from the accretion flow can become important in the midplane of the inner disk \citep{dal06} but we ignore this since our measurements are most sensitive to the outer disk where stellar irradiation dominates. The values of $T_{\rm mid}$, $T_{\rm atm}$ and $Z_q$ vary with radius according to:
\begin{eqnarray}
T_{\rm mid} = T_{\rm mid0}\left(\frac{r}{150 \rm AU}\right)^{q}\nonumber\\
T_{\rm atm} = T_{\rm atm0}\left(\frac{r}{150 \rm AU}\right)^{q}\\
Z_q = Z_{q0}\left(\frac{r}{150 \rm AU}\right)^{1.3}\nonumber.
\end{eqnarray}
This type of profile for the vertical and radial temperature structure is consistent with detailed models of the gas \citep{jon07,wal10} and the dust \citep{dal06}. We only consider the gas temperature in our models and are not concerned with any possible differences between the gas and dust temperature that can arise in the uppermost layers of the disk \citep{jon07}.

At each radius the vertically integrated mass is set by the surface density profile, which we take to follow the form:
\begin{equation}
\Sigma_{\rm gas}(r) = \Sigma_c\left(\frac{r}{R_c}\right)^{-\gamma}\exp\left[-\left(\frac{r}{R_c}\right)^{2-\gamma}\right].
\end{equation}
The value of $\Sigma_c$ depends on the total gas mass, the radial power law index of the surface density and the critical radius ($\Sigma_c=M_{\rm gas}(2-\gamma)/(2\pi R_c^2)$). Unless otherwise specified we assume $M_{\rm gas}$=0.09M$_{\odot}$ \citep{ise07}. The critical radius $R_c$ defines the turnover between a power law and a sharp exponential decay. This is motivated theoretically by a similarity solution to the disk structure calculation for a disk with viscosity that scales with radius ($\nu \propto r^{\gamma}$) \citep{lyn74,har98}. \citet{and10} use this surface density profile to model resolved sub-mm dust emission from a collection of young solar-type stars and find $\gamma$ typically close to 1 with $R_c$ ranging from 15 to 200 AU. This functional form has also been successful in fitting both the dust and the gas \citep{hug08} including the different apparent radii for the gas and dust disk, although recent evidence suggests that in some systems the dust disks are smaller than the gaseous component \citep{and12,deg13,wal14}. 

In a disk with an isothermal vertical profile, the pressure scale height H is simply defined as the width of the Gaussian vertical density profile. With our smoothly varying temperature profile, the definition of H is not as straightforward, although it is still useful to estimate H for comparison with literature models. When used we define H as $c_s/\Omega$, where the midplane temperature sets the sound speed.

The velocity field is taken to be Keplerian, with modifications due to the height above the midplane and the pressure support of the gas:
\begin{equation}
\frac{v^2}{r} = \frac{GM_*r}{(r^2+z^2)^{3/2}} + \frac{1}{\rho_{\rm gas}}\frac{\partial P_{\rm gas}}{\partial r},
\end{equation}
where $\rho_{\rm gas}$ and $P_{\rm gas}$ and the gas mass density and pressure respectively. \citet{ros13} estimate that these modifications change the velocities by a few percent in the upper layers of the disk. We do not include the self-gravity of the disk since this has a negligible effect on the velocity field, especially for our small gas mass \citep{ros13}. 

Gas-phase CO is present in the regions of the disk where it is not frozen out onto dust grains, or photo-dissociated by high energy stellar photons. In modeling SMA observations of CO emission from the HD 163296 system, \citet{qi11} find a freeze out temperature of 19 K, which we take as the freeze-out boundary in our models. We simulate freeze-out by decreasing the CO abundance by eight orders of magnitude in regions where the gas temperature drops below 19 K.

Photo-dissociation occurs above a height where the vertically integrated column density is  
\begin{equation}
 0.706 \int_{z_{phot}}^{\infty}n_{\rm gas}(r,z)dz  < \sigma_s.
\end{equation}
 This is the same boundary used in \citet{qi11} with $\sigma_s$=0.79$\times$1.59$\times$10$^{21}$cm$^{-2}$. This boundary is defined by the absorption of UV photons from the central source \citep{aik99} and is included in our model as a decrease in the abundance of CO by eight orders of magnitude. The exact location of the photo-dissociation boundary may vary between different isotopologues due to selective self-shielding \citep{vis09,mio14}. When modeling $^{13}$CO and C$^{18}$O emission we initially assume ISM values of $^{12}$C/$^{13}$C = 69 and $^{16}$O/$^{18}$O = 557 \citep{wil99}, although we also consider models in which these isotope abundances, as well as the CO/H$_2$ abundance, are allowed to vary. Unless otherwise specified we assume CO/H$_2$=10$^{-4}$. Our simple chemical model for CO ignores much of the complexity of disk chemistry \citep{wal10,jon07} but has proven successful in fitting the CO emission from the HD 163296 system in the past \citep{ros13,qi11}.

\subsection{Radiative Transfer}
Once the structure of the disk is set, we calculate the flux through the disk. The line intensity is:
\begin{equation}
I_{\nu} = \int^{\infty}_0S_{\nu}(s)\exp[-\tau_{\nu}(s)]K_{\nu}(s)ds,
\end{equation}
where $s$ is the linear coordinate along the line of sight increasing outward from the observer. The optical depth is $\tau_{\nu}(s)=\int_0^sK_{\nu}(s')ds'$ where $K_{\nu}(s)$ is the absorption coefficient and $S_{\nu}(s)$ is the source function. For simplicity we assume the system is in Local Thermodynamic Equilibrium (LTE) with the level populations given by the Boltzmann equation and the local gas temperature and the source function is approximated by the Planck function. The CO emitting region is at a density that is much higher than the critical density ($\sim10^4{\rm cm}^{-3}$), allowing us to assume LTE without much loss of accuracy \citep{pav07}

The line is assumed to be a Gaussian with width
\begin{equation}
\Delta V = \sqrt{\left(2k_BT(r,z)/m_{CO}\right) + v_{\rm turb}^2},
\end{equation}
and a FWHM of 2$\sqrt{\log2}$$\Delta V$. Throughout much of our modeling we scale $v_{\rm turb}$ to the local sound speed (e.g. $\Delta V = \sqrt{(2k_BT(r,z)/m_{CO}) + (v_{\rm turb}/c_s)^2}$). The sound speed, which is a factor of $\sqrt{m_{CO}/(\mu m_{H})}\approx$3.4 larger than the thermal broadening of CO, is expected to set the velocity scale for turbulence \citep{bal98} and this parameterization allows turbulence to slowly vary with radius and height within the disk. Since we cannot resolve the spatial scale of the turbulence, we do not detect the bulk motion, $v_{\rm turb}$, associated with the eddies but instead measure the RMS of the velocity distribution, $\delta v_{\rm turb}$. For a Maxwell-Boltzmann distribution of velocities, which is expected from numerical simulations of turbulence \citep{sim15}, the RMS and mean velocity are the same to within factors close to unity. Given the similarity of these factors, and the prevalence of the notation $v_{\rm turb}$ in the literature, we use $v_{\rm turb}$ to denote turbulence even though strictly speaking we are measuring the RMS width, $\delta v_{\rm turb}$, of the velocity distribution. 

Finally, we keep the distance (d=122pc), stellar mass ($M_*$=2.3$M_{\odot}$) and position angle (PA=312$^{o}$) fixed. The offset of the disk from the phase center and velocity center of the spectrum do not strongly depend on the other model parameters, and were adjusted to minimize residuals using initial model fits. 

\subsection{Model Fitting\label{model_fitting}}
Using the model described above, we can generate synthetic images for a given set of parameters. We use the MIRIAD task UVMODEL to generate model visibilities sampled at the same spatial frequencies as the data. Even though the data contain both XX and YY polarizations, we only use the total intensity (I=(XX+YY)/2) to compare with the model because the line emission is not expected to be polarized. The goodness of fit between the model (V$_{\rm mod}$) and observed (V$_{\rm obs}$) visibilities is calculated using the chi-squared statistic where the uncertainty at each uv point is derived from the dispersion in the data, as described earlier.

To sample the posterior probability distribution for each of the parameters we use a MCMC technique. In particular, we employ the affine-invariant routine EMCEE \citep{for13} based on the algorithm originally presented in \citet{goo10}. As opposed to a Metropolis-Hastings chain, the affine-invariant method uses a large number of walkers whose movements through parameter space are proposed along lines to the position of the other walkers. This has the advantage of requiring less fine-tuning to accurately sample the posterior distribution function (PDF) and can efficiently probe the PDF even when there are strong degeneracies between parameters, as is sometimes the case with our models. After an initial burn-in period the positions of the walkers along the chains represent independent samples from the PDF and the distribution of the walkers in parameters space can be used as an estimate of the PDF. Our chains typically consist of 40 walkers and 3000 steps. We only examine the last 1000-2000 steps in each chain, corresponding to $>$10 autocorrelation times, resulting in 40000-80000 samples of the PDF. The initial positions of the walkers are chosen to be evenly distributed throughout parameter space. We require that $M_{\rm gas}$, $R_c$, and $v_{\rm turb}$ are positive with uniform priors across the parameter space.

Systematic uncertainties in the amplitude calibration of the data can have a large effect on the results. ALMA line fluxes differ from those previously measured by the SMA for these same lines \citep{qi11,hug11} by $\sim$5\%, which is within the expected 20\%\ limit on the systematic uncertainty of the ALMA and SMA data. Since CO(3-2) is optically thick its flux is directly proportional to the temperature of the $\tau$=1 surface and a systematic uncertainty in the total line flux will directly translate into an uncertainty in the temperature. Since both temperature and turbulence contribute to broadening of the line, an uncertainty in one will translate into an uncertainty in the other (although, as discussed below and in \citet{sim15}, the spatial resolution of the images and the peak-to-trough ratio of the line profile help to break some of this degeneracy). For the optically thin lines whose flux is proportional to both the surface density and the temperature, the exact influence of the systematic uncertainty is less obvious. This is especially important when combining information from multiple lines, which are subject to different systematic uncertainties. 

To account for the systematic uncertainty, we scale all of the model visibilities by a factor of $s$. This factor is defined such that a value larger than one corresponds to the true disk flux being brighter than the data (e.g. $s=1.2$ implies the true flux is 20\%\ brighter than the measured flux), while a value less than one corresponds to a true disk flux weaker than the measured flux. For much of our modeling this is kept fixed at $s$=1, but for CO(3-2) we consider fitting when $s$ is fixed at 0.8 or 1.2, and we treat it as a free parameter when fitting the four lines simultaneously. Despite the large number of degrees of freedom, $\sim$millions for the single line fits, including too many free parameters prevents the walkers from converging on a best fit model, especially when there are strong degeneracies involved. For this reason we fix $s$ in the single line fits. In the multi-line fit there is enough additional information to allow for a constraint on $s$, and there we consider situations where $s$ is left as a free parameter. The total number of free parameters depends on the line being fit and ranges from four ($v_{\rm turb}$, $\gamma$, inclination, N(X)/N($^{12}$C$^{16}$O)) when fitting $^{13}$CO and C$^{18}$O to 10 ($q$, $R_c$, $v_{\rm turb}$, $T_{\rm atm0}$, $\gamma$, $T_{\rm mid0}$, inclination, log(CO/H$_2$), $s_{32}$ and $s_{21}$) when fitting all four emissions lines simultaneously while accounting for the systematic uncertainty.

\section{Results\label{results}}

\subsection{CO(3-2)}
We start by modeling the high-resolution CO(3-2) line using 40 walkers over 1500 steps. The first 1000 steps are excluded to remove the burn-in phase of the MCMC chains; with typical auto-correlation times of 20-80 steps this corresponds to when the walkers have settled around the best fit. Since CO(3-2) is optically thick it is not sensitive to the surface density, and we fix $M_{\rm gas}$=0.09M$_{\odot}$ and $\gamma$=1, consistent with previous models \citep{ise07,qi11}. We allow $R_c$ to vary to control the radial extent of the emission. By spatially resolving the front ($z>0$) and back ($z<0$) side of the disk, along with the cold midplane, we effectively have two data points to confine T(z), which is not enough to fully constrain the three parameters that control T(z) ($T_{mid0}$, $T_{atm0}$, and $Z_{q0}$). Our initial attempts at MCMC modeling found that $T_{\rm atm0}$ and $Z_{q0}$ were highly degenerate, which slowed significantly the convergence of the walkers. Given the preference of the models for higher $Z_{q0}$ we fix $Z_{q0}$=70AU. With many resolution elements across the radius of the disk, we can constrain the radial profile of the temperature, and allow q to vary. Finally, we include turbulence as well as inclination.

\begin{figure}
\center
\includegraphics[scale=.4]{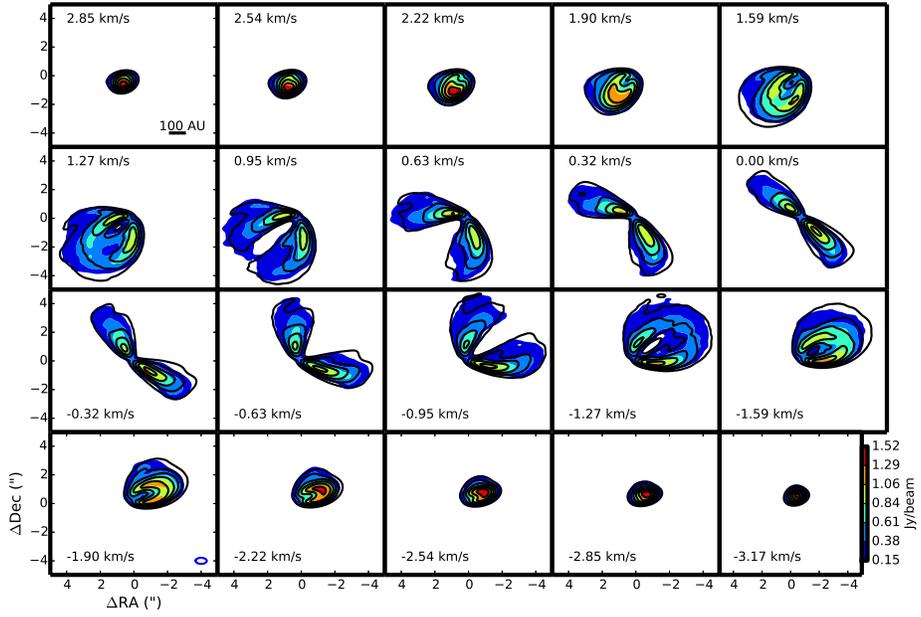}
\caption{Images of select channels from the high-resolution CO(3-2) data (colored filled contours) along with the best fit model (black contours). Contour levels are set at 10\%,25\%,40\%,... of the peak flux, where 10\%\ peak flux=0.15Jy/beam$\sim$18$\sigma$, as marked on the scale. Overall the model successfully reproduces much of the emission.\label{chmaps_co32}}
\end{figure}

\begin{figure}
\center
\includegraphics[scale=.4]{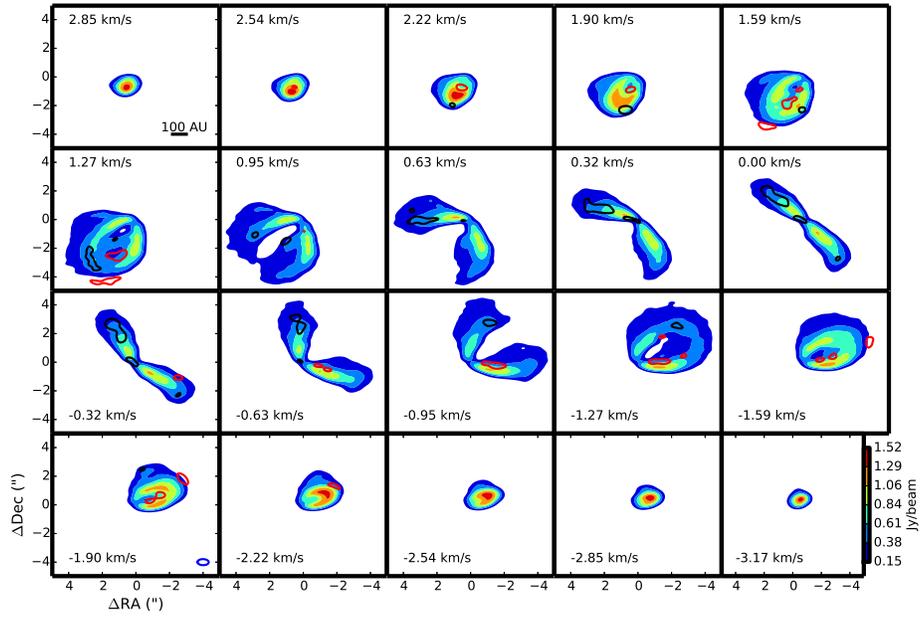}
\caption{Images of select channels from the high-resolution CO(3-2) data (colored filled contours) along with images generated from the difference in visibilities (red and black contours at 10\%,25\%,40\%,... of the peak flux). Red contours mark where the model is brighter than the data while black contours mark where the data is brighter than the model. The model does match the overall shape and strength of the emission, while there are still some residuals.\label{imdiff_co32}}
\end{figure}

\begin{figure}
\includegraphics[scale=.5]{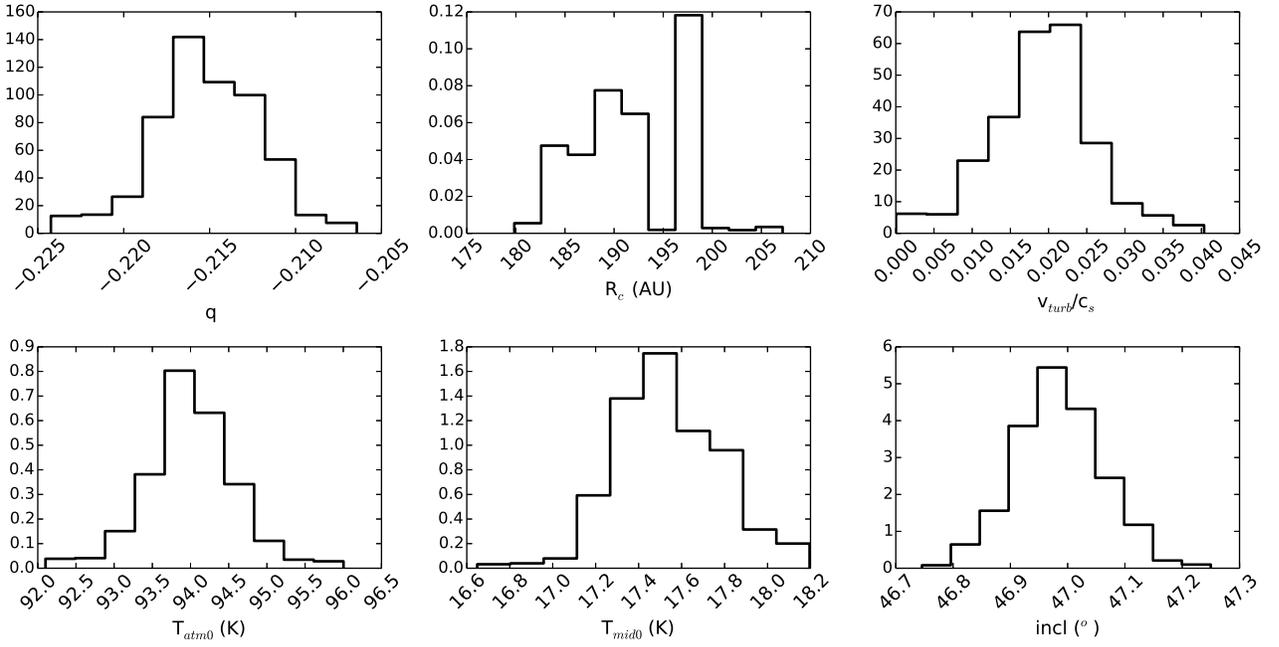}
\caption{PDFs, normalized so that the total area under each distribution is one, for each parameter derived from fitting the low-resolution CO(3-2) data.\label{pdfs_co32}}
\end{figure}

Results are listed in Table~\ref{standard}, along with the three-sigma uncertainties about the median. Channel maps for the best-fit model are shown in Figure~\ref{chmaps_co32}. We also show residual channel maps (subtracted in the visibility domain, ie. $\Delta I_V=(V_{\rm obs}(u,v)-V_{\rm mod}(u,v))$) in Figure~\ref{imdiff_co32}. This model reproduces much of the morphology of the disk, including the front/back emission with a drop in emission at the cold midplane, and has a reduced chi-squared close to one. There are still some residuals between the model and the data but these may be due to deviations from our simple functional form for the temperature and density structure, or even non-circular motion, that will not be perfectly captured in our model. A fit directly to the low-resolution data returns a nearly identical, although slightly weaker limit on turbulence (Table~\ref{standard},Figure~\ref{pdfs_co32}), while also producing a good fit to the data. Slight variations in the temperature structure are consistent with degeneracies and are well below the systematic uncertainties, discussed below. The similarity in the fit to these spectra with different velocity resolutions indicates that the spatial information is providing much of the constraint on the models. Given the consistency between the high and low-resolution data results, while we report the structure from the high-resolution PDFs, we explore various aspects of the fitting using the low-resolution spectra.

\begin{figure}
\center
\includegraphics[scale=.4]{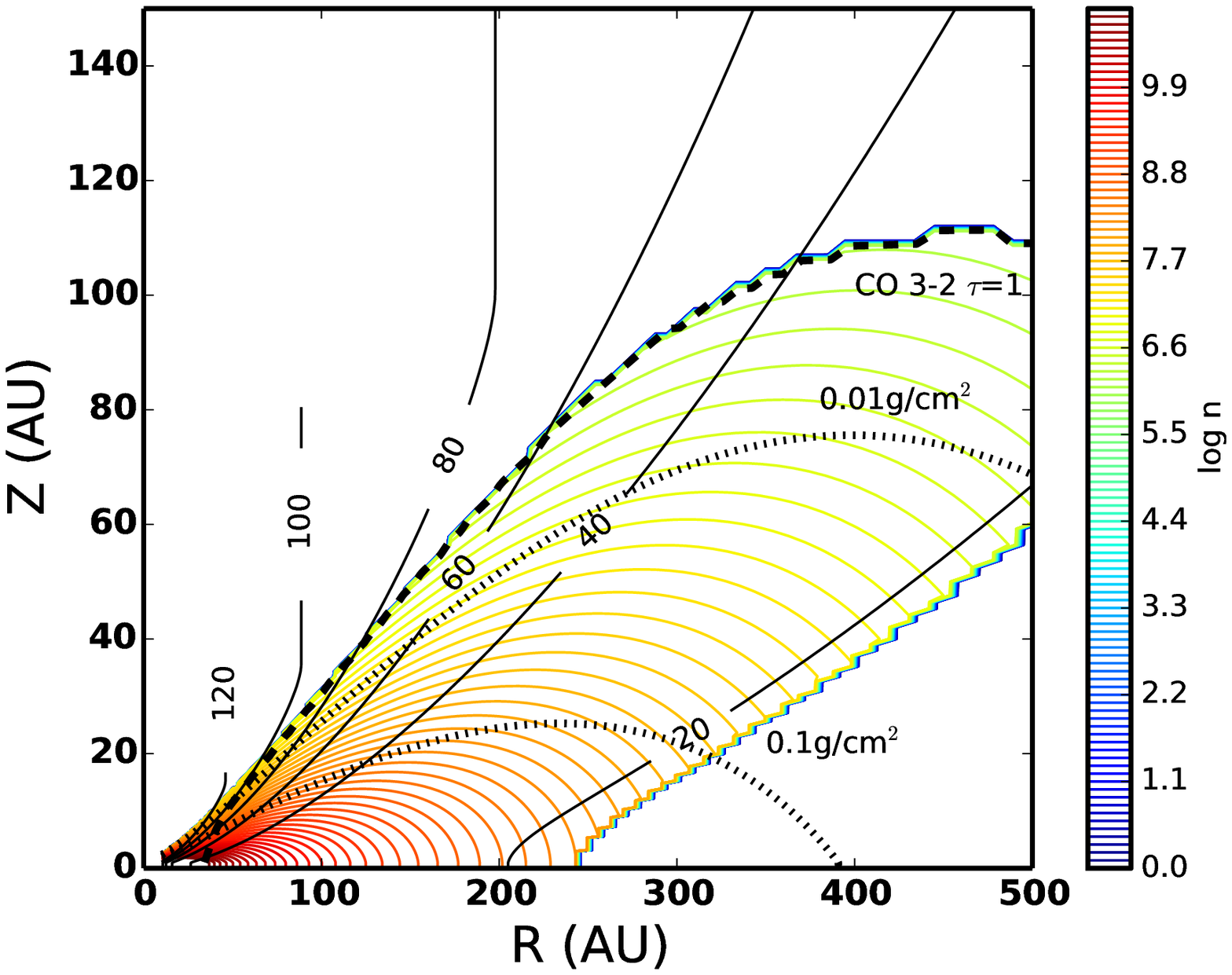}
\includegraphics[scale=.4]{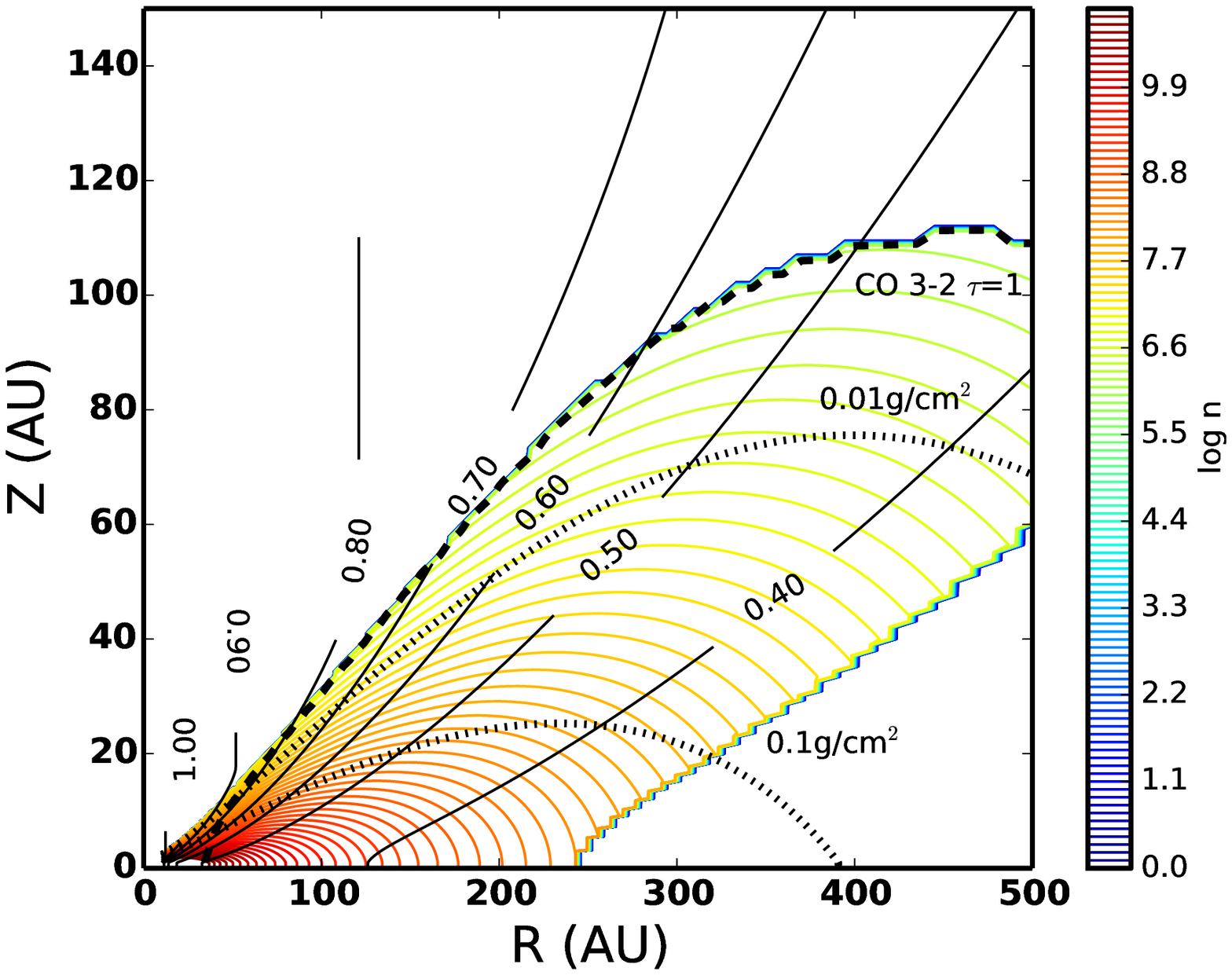}
\caption{Gas density (colored contours) for the best model fit to the low-resolution CO(3-2) data. The density is only marked in the region of the disk where CO exists. In the left panel the black contours show the gas temperature while in the right panel the black contours mark the sound speed in units of km s$^{-1}$. In both panels the dashed line shows the location of the CO(3-2) $\tau$=1 surface, which effectively traces the photodissociation boundary, while the upper/lower dotted lines mark the height where the cumulative surface density, as measured from z=170AU, is equal to 0.01 and 0.1 g cm$^{-2}$.\label{disk_struct_co32}}
\end{figure}

The temperature and density structure for the best fit model are shown in Figure~\ref{disk_struct_co32}. The CO is confined to a thin layer that, at 200 AU, stretches from 10 to 60AU above the midplane with temperatures ranging from 19 K to $\sim$70 K. Previous models \citep{qi11,til12,deg13} of the disk around HD 163296 employed radiative transfer calculations of the equilibrium disk temperature, assuming a dust distribution and a luminosity for the central source, and found similar temperatures to those in our models. This is not surprising since CO(3-2) is optically thick and strongly sensitive to the temperature. Our radial temperature profile falls off with a power law index of -0.278$^{+0.003}_{-0.008}$ (-0.216$^{+0.01}_{-0.008}$ for the low-resolution fit), significantly shallower than the canonical value of -0.5, but consistent with values expected from radiative transfer models that take into account the flaring of the outer disk \citep{dal06}. Flaring leads to more direct interception of starlight in the outer disk, thereby slowing the rate at which the temperature falls off with distance from the central star. 

While the statistical errors on the model parameters are small, they do not account for the systematic effects. For example, the uncertainty in distance strongly influences $q$ since a more distant disk requires more extended radial emission which can be accomplished with a flatter radial temperature profile. Adjusting the distance by $\pm$10pc, its one-sigma uncertainty \citep{mon09}, and rerunning the fit to the low-resolution data we find that $q$ varies by up to 20\%\ ; $q$ rises from -0.236$\pm0.004$ to -0.175$^{+0.01}_{-0.02}$ as distance increases, compared to -0.216$^{+0.01}_{-0.008}$ at the fiducial distance. The critical radius also increases with distance, from 158$^{+6}_{-5}$ to 203.7$^{+19}_{-5}$ AU for the near and far distances respectively, while the other model parameters are consistent within the statistical uncertainty. The differences between $q$ and $R_c$ when fitting the high-resolution and low-resolution data (Table~\ref{standard}) also highlight the degeneracy between these two parameters. The maximum spatial extent of the CO emission can be matched with a narrow disk (low $R_c$) and shallow temperature profile (high $q$) or a wide disk (high $R_c$) and a steeper temperature profile (low $q$). This effect is small, leading to 10-20\%\ differences in $q$ and $R_c$, but is large compared to the statistical uncertainties. 

Similarly, the statistical uncertainty on the atmosphere temperature is $<$5\% although this does not include the uncertainty on the amplitude calibration. As discussed earlier, we account for the systematic uncertainty by applying a scale factor to the model and we run additional MCMC trials fitting the low-resolution data with $s$ fixed at 1.2 and 0.8, mimicking a true disk flux that is 20\%\ higher/lower than the data. The biggest effect is in $T_{\rm atm0}$ which varies from 84.3$^{+0.7}_{-2.0}$ to 101$\pm3$ for low/high disk flux respectively, compared to 94$\pm2$ in the fiducial low-resolution model fit. The influence of systematic uncertainty dwarfs the statistical uncertainties for $T_{\rm atm0}$. The variations of the other parameters in response to the uncertainty in the flux calibration are within the statistical errors derived from the PDF.

\begin{figure}
\includegraphics[scale=.4]{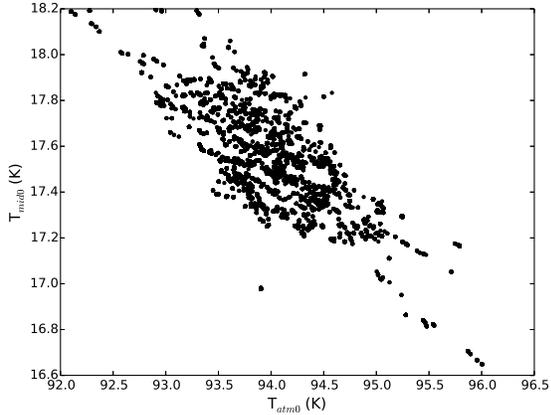}
\caption{In fitting the low-resolution CO(3-2) data we see evidence for a degeneracy between T$_{\rm mid0}$ and T$_{\rm atm0}$. As T$_{\rm mid0}$ increases the disk becomes puffier, raising the $\tau=1$ surface higher in the disk where it is warmer, and T$_{\rm atm0}$ must then adjust downward to maintain the same flux from the $\tau=1$ surface.\label{degeneracies_co32}}
\end{figure}

The atmosphere temperature is also subject to its degeneracy with the midplane temperature (Figure~\ref{degeneracies_co32}). While the emission from the bright upper half of the disk constrains the atmosphere temperature, the midplane temperature is constrained through the emission from the faint lower half of the disk and the lack of emission between these two features. In our fiducial model we find $T_{\rm mid0}$=21.8$^{+0.7}_{-0.4}$. The constraints on $q$ and $T_{\rm mid0}$ in turn lead to a CO ice line that falls within the range $R_{\rm ice}$=245$^{+27}_{-13}$AU while previous observations of chemical tracers of CO freeze-out find that it is closer to 150-160AU \citep{qi13b,mat13}. Overestimating the midplane temperature results in a more puffed up disk, pushing the $\tau=1$ surface higher in the disk and the atmosphere temperature must be reduced accordingly to maintain the same temperature at the $\tau=1$ surface. This can be seen when looking at the results from fitting the low-resolution spectra which has different values of $T_{\rm mid0}$ and $T_{\rm atm0}$, and predicts a CO ice line at $R_{\rm ice}$=103$^{+23}_{-18}$AU, but produces similar images. 

While our models do not treat the dust, assumptions about the dust distribution may still be implicit in our models. \citet{jon07} find that in their chemical models the depletion of dust from the upper layers of the dust leads to a drop in the gas temperature. In the context of the \citet{dal06} models, \citet{qi11} show that the settling of large dust grains will affect the height at which the disk reaches the atmosphere temperature; in other words more settling results in lower $Z_{q0}$. Our assumed value of $Z_{q0}$ is most similar to models with little dust settling \citep{qi11}.
 The significant degeneracy between $Z_{q0}$ and $T_{\rm atm0}$ means that a model with lower $Z_{q0}$, when combined with a lower $T_{\rm atm0}$, would fit the data as well as the models presented here. \citet{ros13}, using the same functional form for the vertical temperature profile as we do here, find that the data can be fit with $T_{\rm atm0}$=64K, given $Z_{q0}$=43 AU. This atmosphere temperature is much lower than our $T_{\rm atm0}$=86$\pm1$K but the difference in $Z_{q0}$ means that both models produce a nearly identical gas temperature at the $\tau=1$ surface. Further data probing the uppermost layers in the disk atmosphere, where CO is photodissociated, are needed to break this degeneracy. 

Overall we are able to find a model that accurately fits the data and is consistent with previous radiative transfer models of the HD 163296 system. By carefully accounting for parameter degeneracies and systematic effects due to assumptions about the distance and flux calibration, we can measure the statistical and systematic uncertainties in temperature and density structure of the disk. This indicates that our derived disk structure is not going to strongly bias our characterization of turbulence.

\subsubsection{Turbulence}
\begin{figure}
\includegraphics[scale=.4]{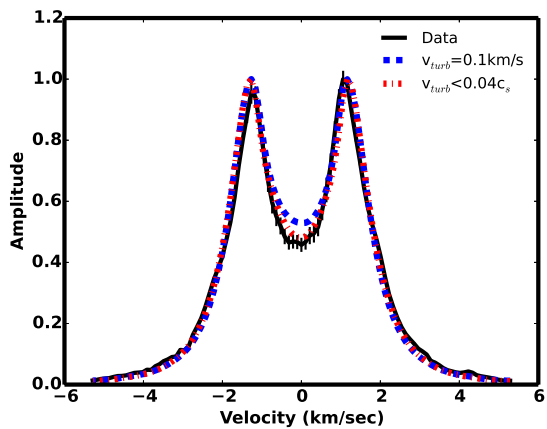}
\caption{CO(3-2) high resolution spectra (black line) compared to the median model when turbulence is allowed to move toward very low values (red dot-dashed lines) or when it is fixed at 0.1 km s$^{-1}$ (blue dashed lines). All spectra have been normalized to their peak flux to better highlight the change in shape. The models with weak turbulence provide a significantly better fit to the data despite the fact that the turbulence is smaller than the spectral resolution of the data.\label{imspec_test}}
\end{figure}

When fitting either the high-resolution or low-resolution CO(3-2) spectra, we consistently find low levels of turbulence (Table~\ref{standard}). We conservatively quote three-sigma upper limits of $v_{\rm turb}<$0.03c$_s$ and $v_{\rm turb}<$0.04c$_s$ respectively. In the uppermost layers of the CO region, these limits correspond to velocity dispersions of less than $\sim$9-19m s$^{-1}$, with higher velocities at smaller radii. While we have some constraint on the inner disk from the high velocity channels, most of the information about the turbulence comes from the spatially resolved outer disk ($R>$30AU) and our upper limit is indicative of the behavior in this region. Our limit is well below the spectral resolution of even the high-resolution data ($\delta$v=0.1km s$^{-1}$). To test whether we can distinguish between turbulence at the resolution limit from much weaker non-thermal motion, we run an additional MCMC fit to the low-resolution CO(3-2) data with $v_{\rm turb}$ fixed at 0.1 km s$^{-1}$ while allowing the other model parameters to vary. Figure~\ref{imspec_test} shows the best fit $v_{\rm turb}=0.1$km s$^{-1}$ model along with our fiducial low turbulence model, using a Gaussian fit to the short-baseline visbilities to derive the spectra. The low turbulence models are a much better fit to the spectra, and the chi-squared, which captures the behavior of the full three-dimensional data set, also finds a significantly better fit ($>$10$\sigma$ significance) from the low-turbulence models. This behavior indicates that fitting a model to the full data set has a stronger diagnostic power than the broadening of the line in the spectral domain. \citet{sim15} have suggested the peak-to-trough ratio as a potential diagnostic of turbulence. They found in simulated observations of numerical MRI simulations, with typical $v_{\rm turb}\sim0.1-0.5$c$_s$ at the CO(3-2) emitting surface, that the peak-to-trough ratio decreases as the turbulence increases. This behavior can be seen in Figure~\ref{imspec_test} where the $v_{\rm turb}=0.1$km s$^{-1}$ models have a much lower peak-to-trough ratio than the low-turbulence models and the data, consistent with the conclusion that the turbulence is weaker than 0.1km s$^{-1}$. 

\begin{figure}
\includegraphics[scale=.4]{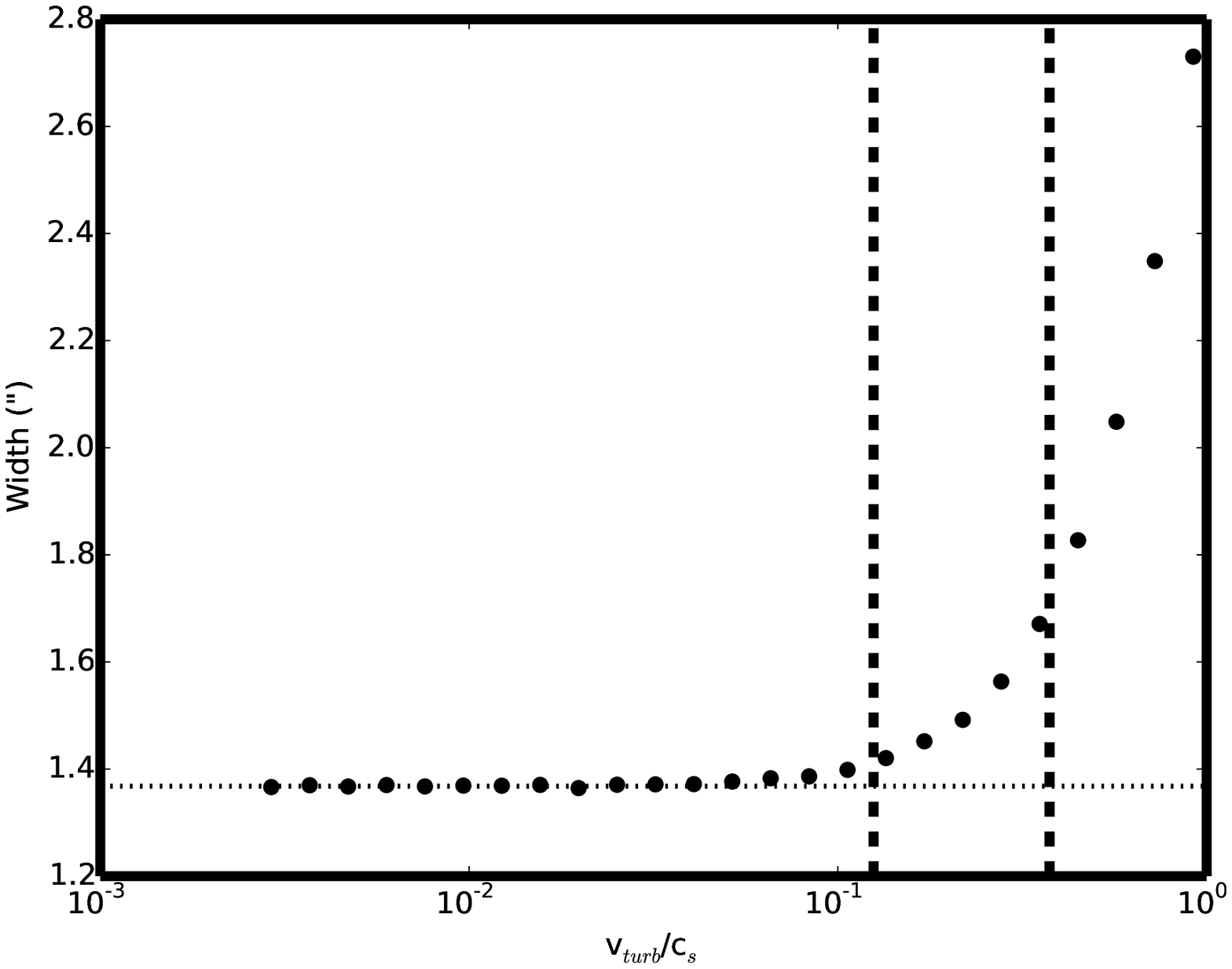}
\includegraphics[scale=.4]{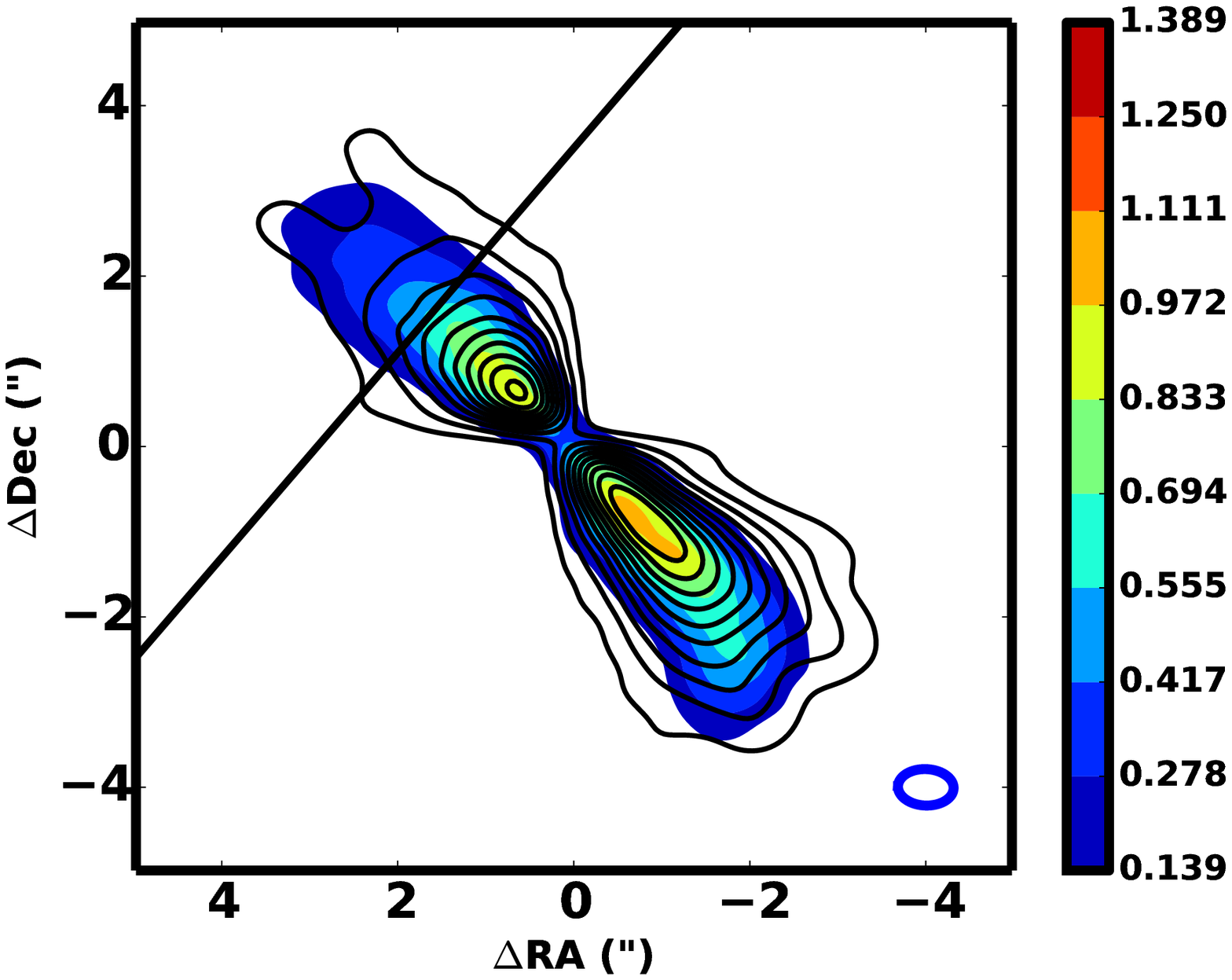}
\includegraphics[scale=1.]{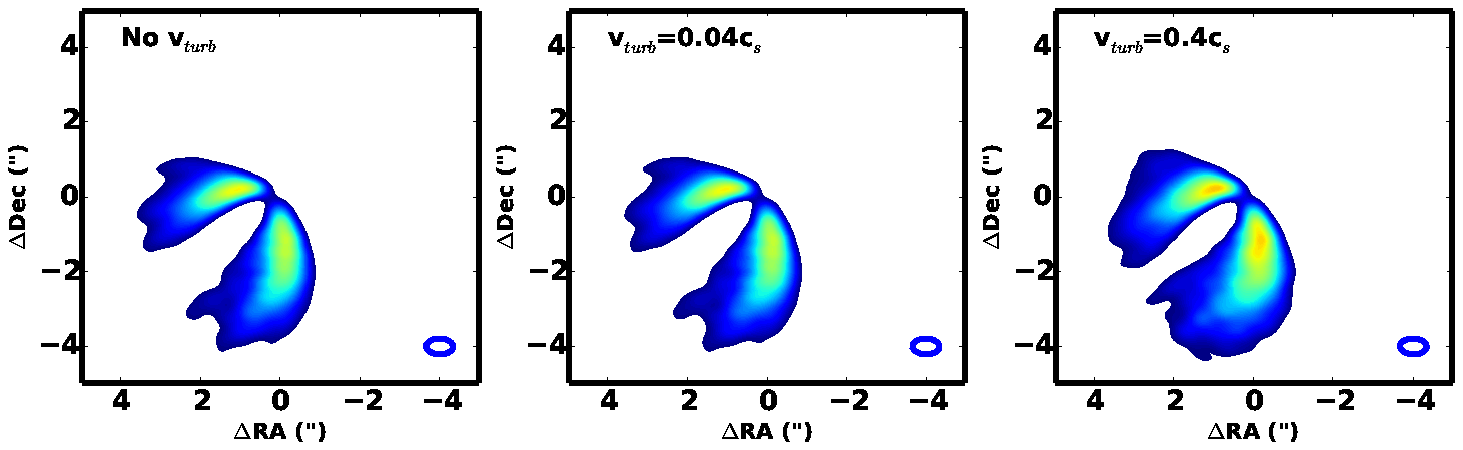}
\caption{[Top] Width of the emission through the central velocity channel (left panel) along a line that runs perpendicular to the emission (marked on the right panel, which shows the data in the central velocity channel along with a v$_{\rm turb}$=0.4c$_s$ model). As turbulence is varied in the models, the emission broadens. The vertical lines mark the spectral resolution for the low-res (right) and high-res (left) CO(3-2) spectra. The images change substantially with turbulence, allowing us to probe non-thermal motion below the spectral resolution of our data. [Bottom] Model channel maps for turbulence varying between zero (left panel), the fiducial model result (center panel) and v$_{\rm turb}$=0.4c$_s$. The flux in these channel maps has been normalized to its peak value to highlight the change in morphology, which is substantial for large turbulence motion.\label{widthvturb}}
\end{figure}

Peak-to-trough ratio is a one-dimensional metric that provides a convenient way of diagnosing differences in the three-dimensional data set. Variations in the temperature can also affect the peak-to-trough ratio, however the uncertainty in temperature is dominated by the uncertainty in the flux calibration and within the range of temperatures given by this uncertainty the peak-to-trough ratio is substantially more sensitive to turbulence than it is to temperature \citep{sim15}. By simultaneously fitting the temperature structure along with the turbulence across the full three-dimensional data set, we account for the various influences of the different parameters on the images as well as the spectral shape. The similarity between the turbulence results when fitting the low-resolution and high-resolution data suggest that the shape of the full line profile is not the dominant factor but instead spatial resolution plays a large role in constraining the non-thermal motion. In the individual channel maps turbulence broadens the width of the disk emission \citep{gui12,sim15}. We demonstrate this effect for the central velocity channel of the HD 163296 system in Figure~\ref{widthvturb} by varying the turbulence while fixing the other model parameters. For low levels of turbulence there is little variation in the images, but the disk broadens dramatically for larger values of turbulence. This occurs even below the spectral resolution of the data, allowing us to push to low levels of turbulence.

The broadening of the images not only changes their shape, but also changes their total flux. This can occur even when the broadening is below the spatial and spectral resolution. In this way line flux could be used as a diagnostic of turbulence, but this leaves it highly degenerate with temperature, and subject to the uncertainty in the calibration of the line flux. The systematic uncertainty is, conservatively, 20\%\ at the ALMA wavelengths and dominates over the channel-to-channel uncertainties. As discussed earlier, we can account for systematic uncertainty by applying a scale factor to the model. When fixing $s=1.2$, i.e. when the true disk flux is 20\%\ brighter than the observed data, we find a limit of $v_{\rm turb}<0.05$c$_s$, while if the true disk emission were 20\%\ fainter than our data the limit on turbulence is $v_{\rm turb}<0.08$c$_s$. Most of the change in disk flux is absorbed by the atmosphere temperature, with only small variations in turbulence. An uncertainty in distance could also affect our result, since the one-sigma uncertainty on distance of 10pc \citep{mon09} leads to a 16\%\ uncertainty in the total flux. In additional MCMC trials with distance fixed at 112 and 132pc we find that the limit on turbulence increases to $v_{\rm turb}<$0.05c$_s$,0.06c$_s$ respectively. Even accounting for the systematic uncertainty and the distance uncertainty, we find that non-thermal motion is limited to very weak levels.

The high-resolution spectrum exhibits an asymmetry in the line profile, with the blue-shifted peak brighter than the red-shifted peak (Figure~\ref{imspec_test}). One possible explanation for the behavior is with an eccentric disk; the disk material 'piles-up' on the slower moving side of the orbit leading to a higher flux from one side of the disk versus the other \citep[e.g.][]{reg11}. Our model includes the dynamics of Keplerian rotation and thermal motion and any additional motion will likely be accounted for by the turbulence parameter. For this reason we choose to report our turbulence results as an upper limit rather than a detection. The posterior distribution for turbulence (Figure~\ref{pdfs_co32}) has a shape characteristic of a detection but given the possibility of unaccounted features being modeled as turbulence, as well as the weak influence of turbulence on the data at such low levels (Figure~\ref{widthvturb}), we conservatively quote an upper limit. Also, as noted earlier, or implementation of turbulence within the models means that we are really constraining the velocity dispersion, or RMS velocity, associated with turbulence rather than the mean velocity within turbulent eddies. 

\subsection{CO(2-1)}
\begin{figure}
\center
\includegraphics[scale=.4]{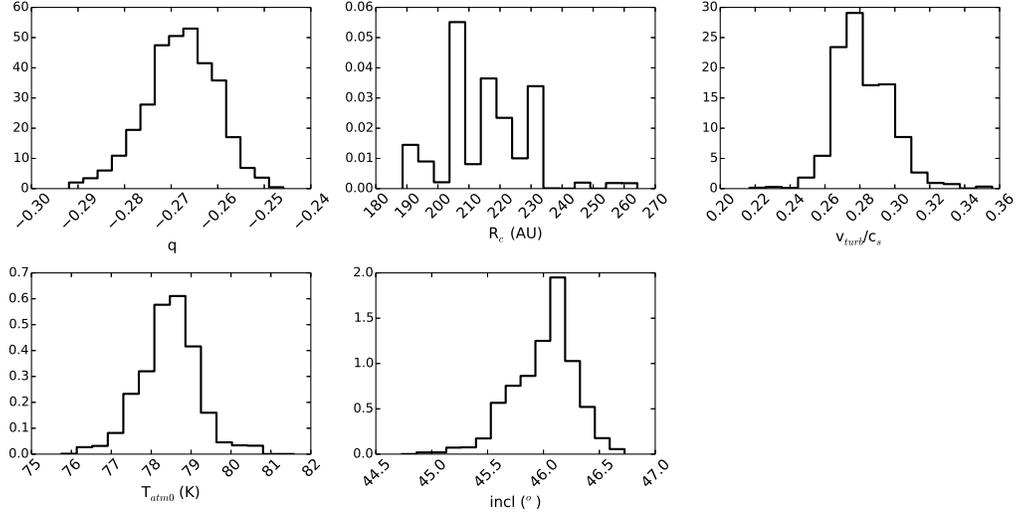}
\caption{Posterior distributions, normalized so that the total area under each distribution is one, for the five parameters used in fitting to the CO(2-1) line.\label{pdfs_co21}}
\end{figure}

\begin{figure}
\center
\includegraphics[scale=.3]{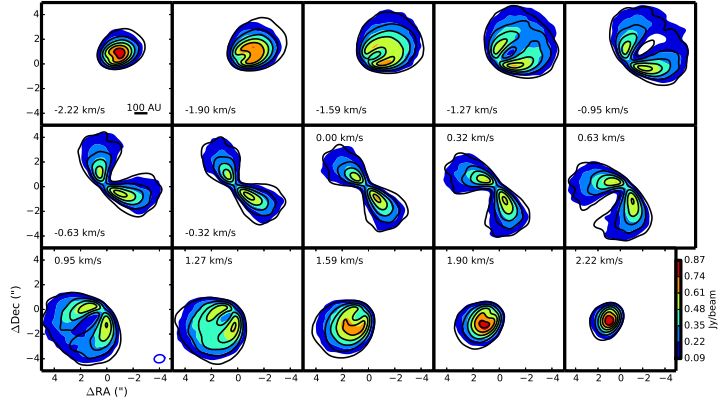}
\caption{Channel maps for the central 15 channels of the CO(2-1) data (colored contours) with the best fit model (black contours). Contour levels are set at 10\%,25\%,40\%,... of the peak flux, where 10\%\ peak flux=0.09Jy/beam$\sim$9$\sigma$.\label{chmaps_co21}}
\end{figure}

\begin{figure}
\center
\includegraphics[scale=.3]{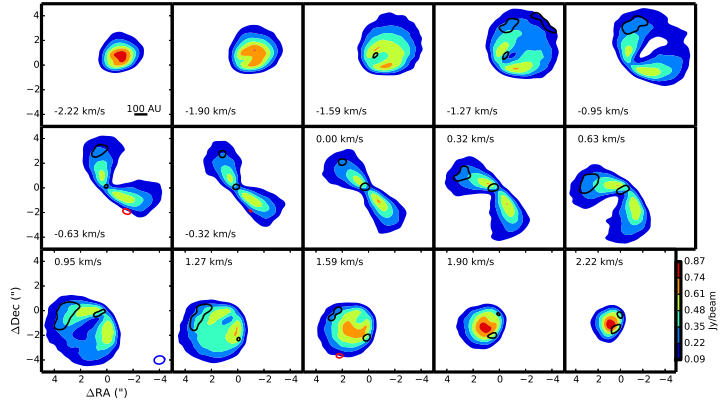}
\caption{Channel maps for the central 15 channels of the CO(2-1) data (colored filled contours) along with images generated from the difference in visibilities (red and black contours). Red contours mark where the model is brighter than the data while black contours mark where the data is brighter than the model. The model matches much of the images.\label{imdiff_co21}}
\end{figure}

As with CO(3-2), our CO(2-1) MCMC trial uses 40 walkers over 3000 steps. With a lower spatial resolution than the CO(3-2) data, the walkers here take longer to converge and the number of steps has been increased accordingly. CO(2-1) is also optically thick, making it sensitive to the temperature structure, but lacks the spatial resolution to distinguish the near and far side of the disk. For this reason we fix $T_{\rm mid0}$=17.5, based on the low-resolution CO(3-2) results, while allowing $q$, $R_c$, $v_{\rm turb}$, $T_{\rm atm0}$, and inclination to vary during the fitting. Results, after removing the first 1000 steps, are listed in Table~\ref{standard} with posterior distributions shown in Figure~\ref{pdfs_co21}. Channel maps and imaged residual visibilities are shown in Figures~\ref{chmaps_co21} and \ref{imdiff_co21} respectively. In the channel maps, outside of minor residuals, we are able to match the radial extent of the disk, as well as the flux throughout the spectra.

Differences in the model parameter PDFs exist between the CO(2-1) and CO(3-2) models, but these are most likely the result of degeneracies between parameters. In the CO(2-1) data we find a strong anti-correlation between $R_c$ and $q$ and compared to the low-resolution CO(3-2) fit, the CO(2-1) line settles on a model with steeper $q$ and larger $R_c$, consistent with this anti-correlation. We also find a smaller inclination and lower $T_{\rm atm0}$. The resulting disk structure derived from the two lines still has very similar temperatures at the top of the CO emitting surface, to which these data are particularly sensitive. 

\begin{figure}
\center
\includegraphics[scale=.3]{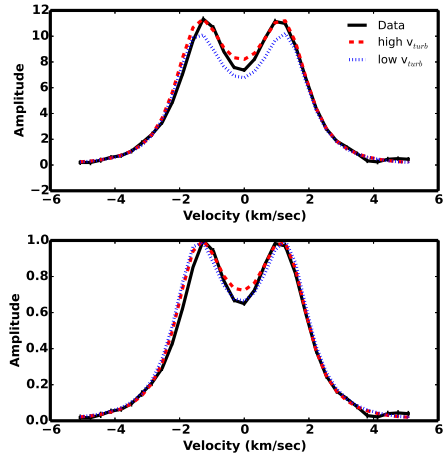}
\caption{Visibility spectra for CO(2-1) comparing the data (solid black line) to a model with low turbulence (v$_{\rm turb}$=0.037c$_s$, blue dotted line) and a model with high turbulence (v$_{\rm turb}$=0.3c$_s$, red dashed line). The other parameters in these models have been allowed to adjust to find the best fit. The bottom panel shows the spectra normalized to their peak flux to better highlight the differences in line shape. While the high turbulence model nominally provides a better fit to the CO(2-1) data, the low turbulence model better matches the line profile, in particular the peak-to-trough ratio, which is sensitive to the turbulence. \label{co21_hiloturb}}
\end{figure}

\begin{figure}
\center
\includegraphics[scale=.3]{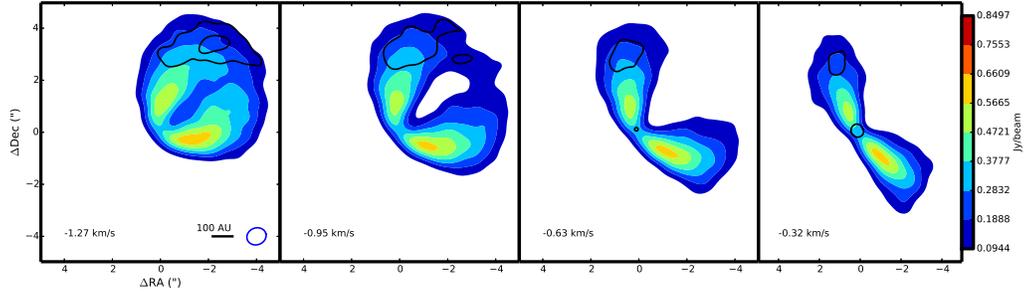}
\caption{Imaged residuals (black solid contours) compared to the CO(2-1) data (colored filled contours) for the low-turbulence fit. The residuals, indicating that the model underestimates the data, are only found on the north-east side of the disk. This asymmetry is likely a radiative transfer effect related to the different heights probed on the near and far side of the disk, and the resulting temperature structure. When the models cannot fully adjust the temperature structure to account for this asymmetry, they increase the turbulence leading to an overestimate of its true value. \label{co21_lowturb_resid}}
\end{figure}

The upper limit on turbulence from CO(2-1) ($v_{\rm turb}<0.31$c$_s$) is substantially higher than derived from the CO(3-2) line, most likely due to the lower spatial resolution of these data compared to CO(3-2). This limit corresponds to velocity widths of 200-150 m s$^{-1}$ in the upper layers of the disk. To see if a satisfactory fit can be found with low turbulence, we run an additional trial with v$_{\rm turb}$ fixed at 0.04c$_s$, ie. the upper limit determined by the low-resolution CO(3-2) data. The resulting spectrum for this model is shown in Figure~\ref{co21_hiloturb}, along with the high turbulence CO(2-1) fit. The low turbulence fit underestimates the overall line flux, but better matches the line shape, in particular the peak to trough ratio. The imaged residuals for the low-turbulence model show features only on the north-east side of the disk (Figure~\ref{co21_lowturb_resid}). This asymmetry may be related to the temperature structure of the disk, as probed by an optically thick tracer. As discussed in \citet{dar03}, the line of sight to an inclined disk probes different heights for the area of the disk closer to the observer, in this case the south-west side, than on the far (north-east) side of the disk. This is because a line of sight through the near side of the disk ends at at smaller radius, where the temperature and densities are higher, than where it entered the disk. Conversely, a line of sight through the far side of the disk has its $\tau$=1 surface in an area with a lower density and temperature than where it entered the disk. While our ray-tracing code will account for photons coming from different surfaces, our assumed functional form for the temperature may not capture the difference in temperature between these surfaces. This would show up as asymmetric residuals, such as those seen in CO(2-1) (Figure~\ref{co21_lowturb_resid}) and in CO(3-2) (Figure~\ref{imdiff_co32}). The MCMC routine cannot adjust the temperature parameters to account for the deviations from the assumed functional form, and tries to remove the residuals instead by increasing the turbulence, which raises the overall flux of the disk. Since the CO(2-1) data has a lower spatial resolution than CO(3-2), it has more room to increase the broadening of the images before the model becomes significantly wider than the data. The end result is a high value of turbulence, and a worse fit to the peak-to-trough ratio, due to limitations of the assumed temperature structure rather than large non-thermal motions within the disk. For this reason we quote an upper limit on turbulence despite the shape of the posterior distribution (Figure~\ref{pdfs_co21})

\subsection{$^{13}$CO(2-1)}
\begin{figure}
\center
\includegraphics[scale=.4]{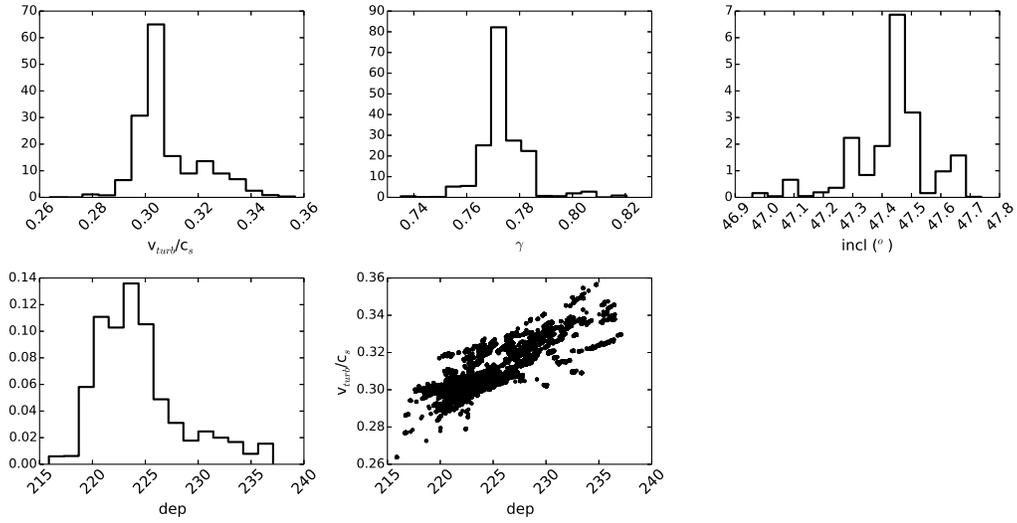}
\caption{Posterior distributions, normalized so that the area under each distribtion is one, for the four parameters used in fitting to the $^{13}$CO(2-1) line. We also find evidence for a degeneracy between turbulence and the depletion factor (=the $^{13}$CO/CO abundance ratio); as the depletion of $^{13}$CO increases more of the emission arises from close to the midplane diminishing the vertical extent of the disk in the images and to match the full spatial extent of the emission the turbulence must be increased. \label{pdfs_13co21}}
\end{figure}

\begin{figure}
\center
\includegraphics[scale=.3]{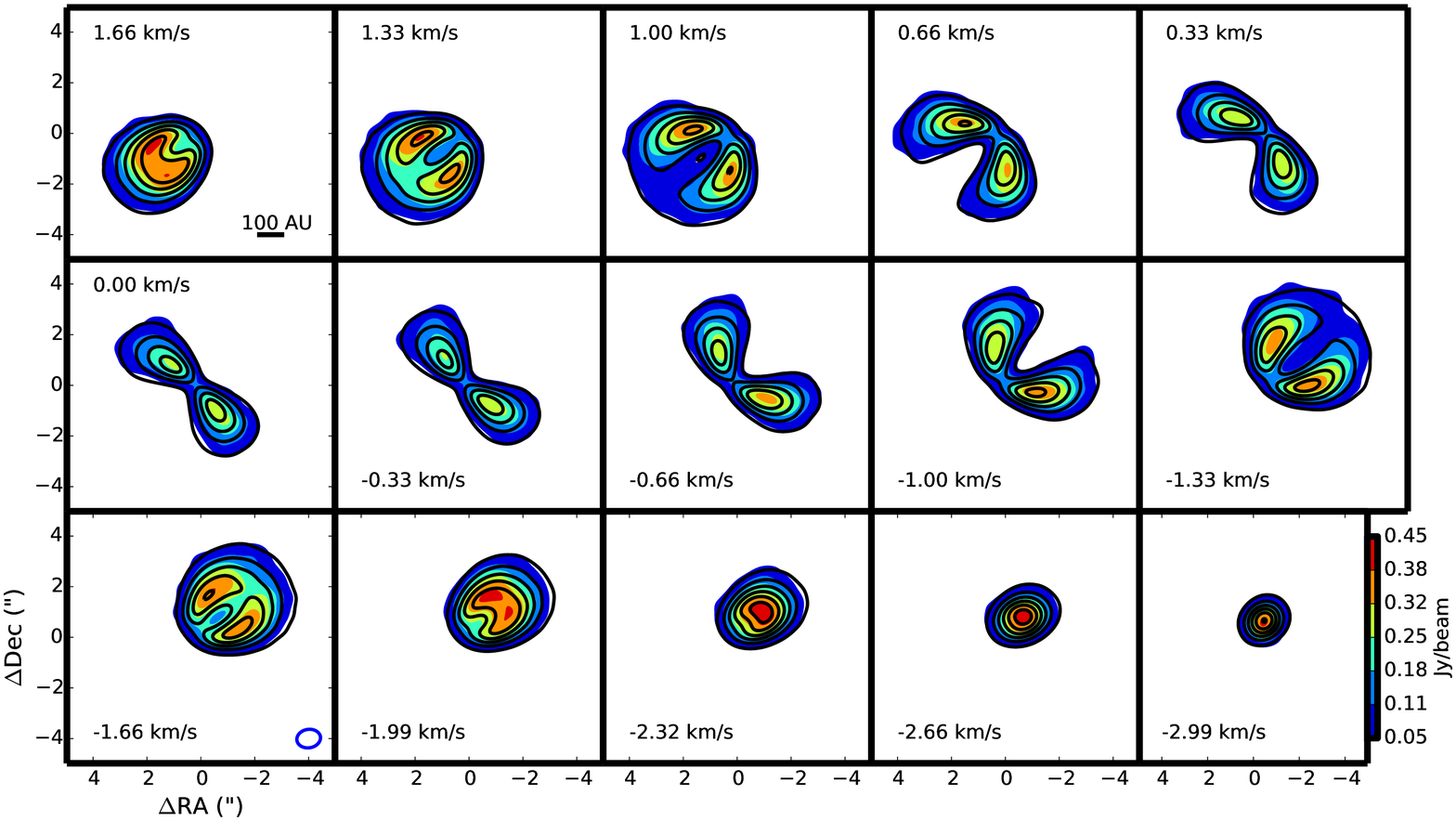}
\includegraphics[scale=.3]{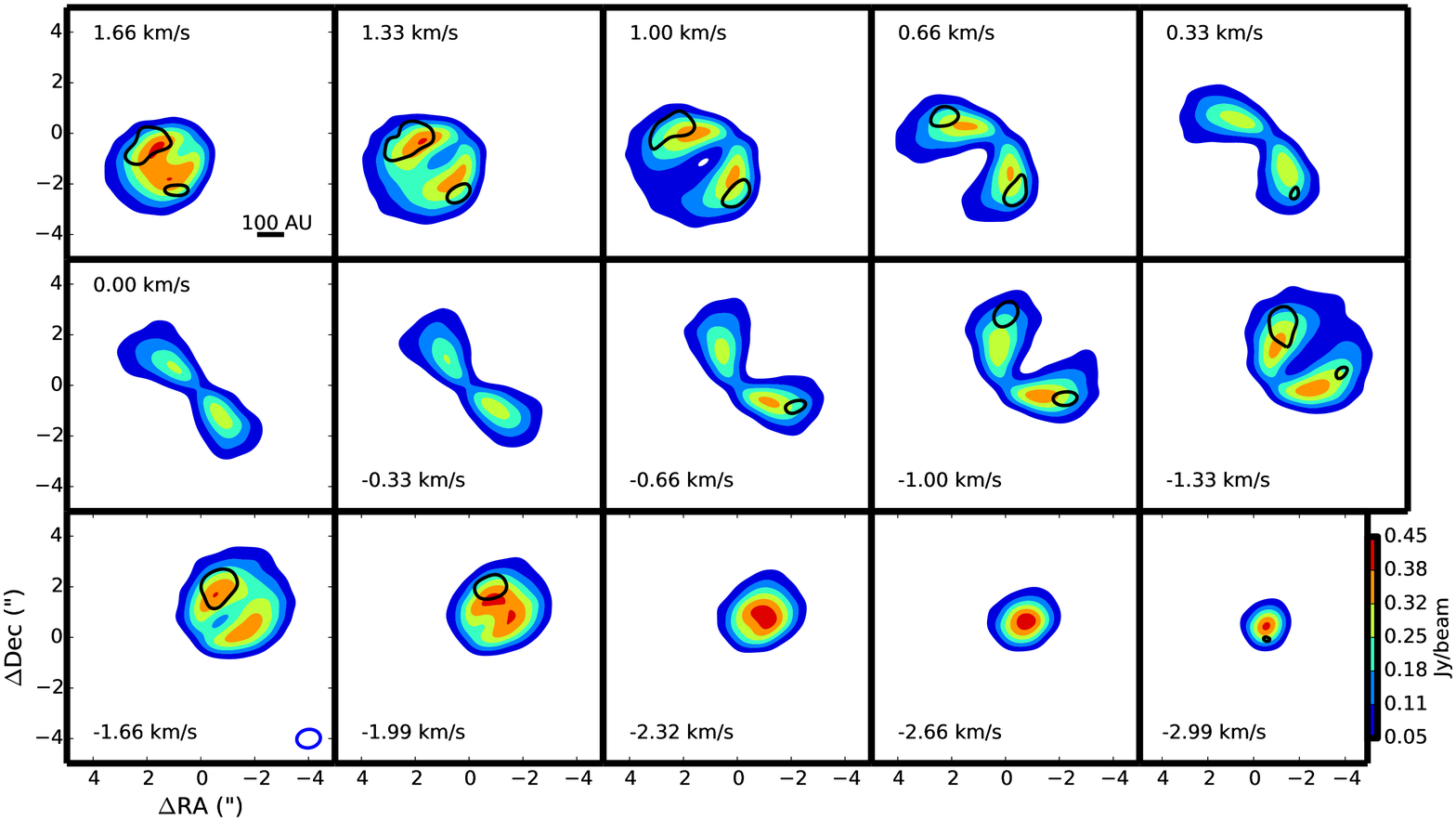}
\caption{Channel maps for the model (top, solid black contour) and the image generated from the difference in visibilities (bottom,red are model$>$data, black are data$>$model) for the fit to the $^{13}$CO(2-1) data (filled contours). Contours are set at 10\%,25\%,40\%,... of the pak flux, where 10\%\ peak flux=0.045 Jy/beam$\sim$8$\sigma$. This model is able to successfully reproduce much of the emission.\label{chmaps_13co21}}
\end{figure}

The decreased abundance of $^{13}$C relative to $^{12}$C reduces the optical depth of $^{13}$CO relative to $^{12}$CO which makes the line emission more sensitive to the total mass of the disk, but introduces a degeneracy between column density and temperature. To avoid this degeneracy we fix the temperature structure, defined by $q$, $T_{\rm atm0}$, and $T_{\rm mid0}$, based on the low-resolution CO(3-2) observations. We then allow $\gamma$ to vary, along with $v_{\rm turb}$ and inclination. We fix the disk mass and $R_c$, while leaving the depletion of $^{13}$CO as a free parameter. During our initial attempts at MCMC modeling, and in the multiline fits discussed below, we found the best possible fit to the data came when the disk mass was fixed and the abundance was allowed to vary. Specifically we treat the $^{13}$CO/CO depletion as a free parameter, although in the case of a single line fit this is equivalent to allowing CO/H$_2$ to vary. Model fits, after removing the burn-in, are listed in Table~\ref{standard} with posterior distributions shown in Figure~\ref{pdfs_13co21}. These models are able to accurately reproduce much of the $^{13}$CO emission, although there are some discrepancies near the line peaks (Figure~\ref{chmaps_13co21}). This is likey due to the temperature structure; the residuals show the same north-south asymmetry as in CO(2-1) and CO(3-2). In the channel maps and imaged residuals (Figure~\ref{chmaps_13co21}) the differences between the model and the data are small, with the strongest residuals only 10\%\ of the peak flux. 

As with CO(2-1), we put a limit on the turbulent broadening ($v_{\rm turb}<$0.34c$_{s}$) that is less strigent than CO(3-2). At the coldest layers close to the midplane, just above CO freeze-out, this corresponds to a non-thermal velocity width less than $\sim$130m s$^{-1}$. We also find $^{13}$CO/CO=233$^{+3}_{-15}$, significantly different from the ISM value of 69 \citep{wil99}. This may be a sign of unaccounted for CO chemistry, such as selective photodissociation \citep{mio14}, or a sign that the assumed disk mass, which is derived from the dust emission, is an overestimate. This result is discussed in more detail below in the context of the multiline fit.

\subsection{C$^{18}$O(2-1)}
\begin{figure}
\center
\includegraphics[scale=.4]{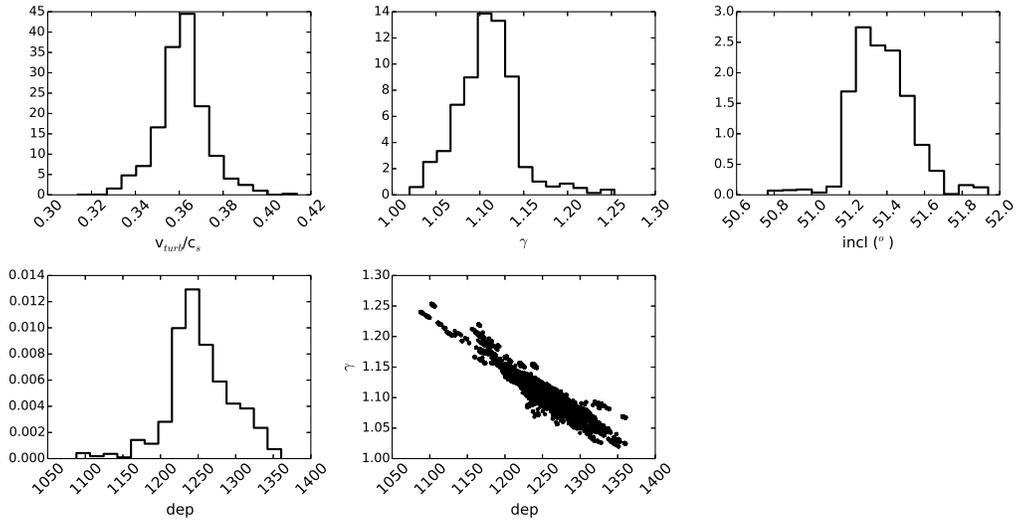}
\caption{Posterior distributions, normalized so that the area under each distribution is one, for the four parameters used in fitting to the C$^{18}$O(2-1) line. We find a strong anticorrelation with the depletion factor and $\gamma$; this is not surprising since both control the surface density of gas. \label{pdfs_c18o21}}
\end{figure}

\begin{figure}
\center
\includegraphics[scale=.3]{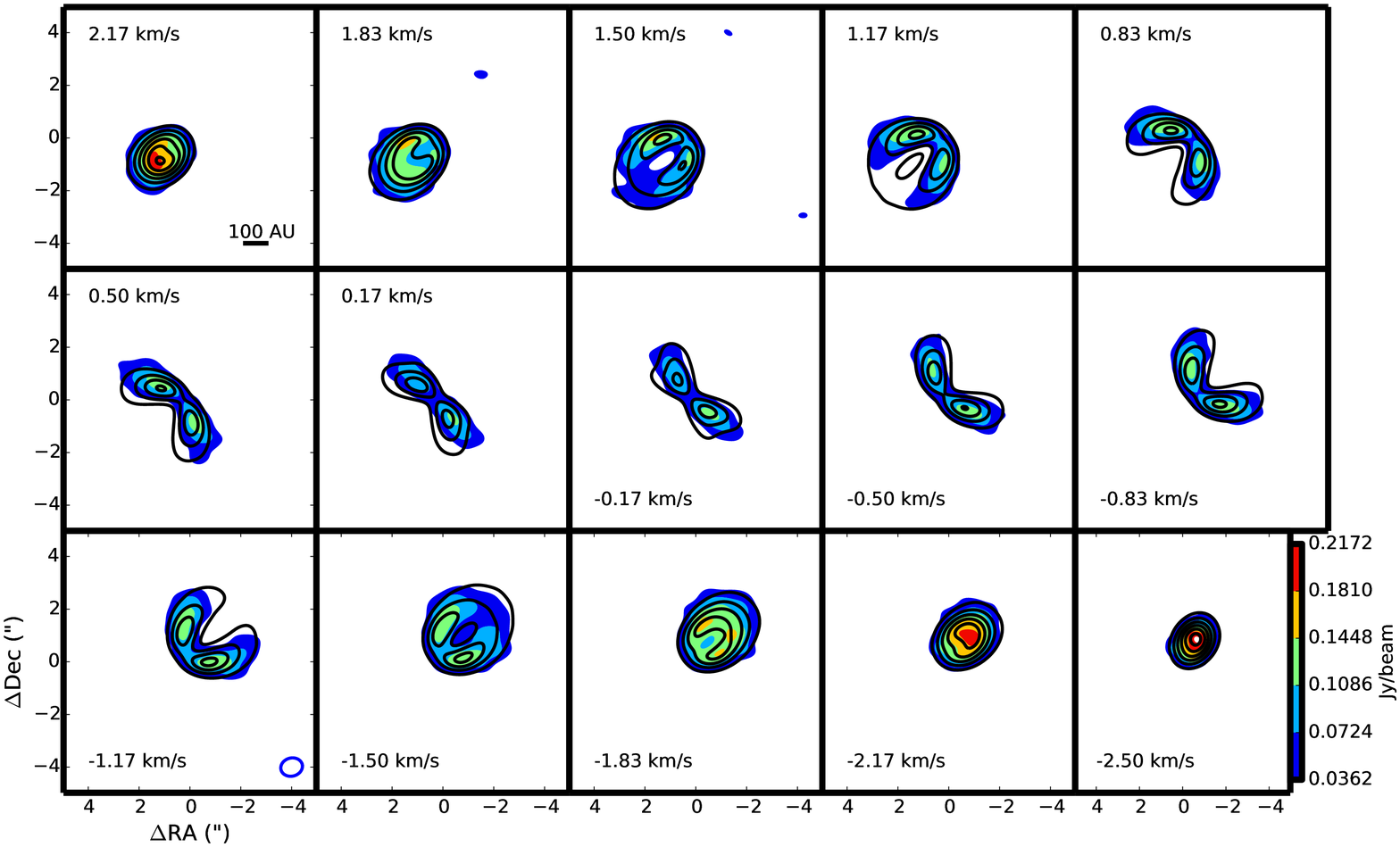}
\includegraphics[scale=.3]{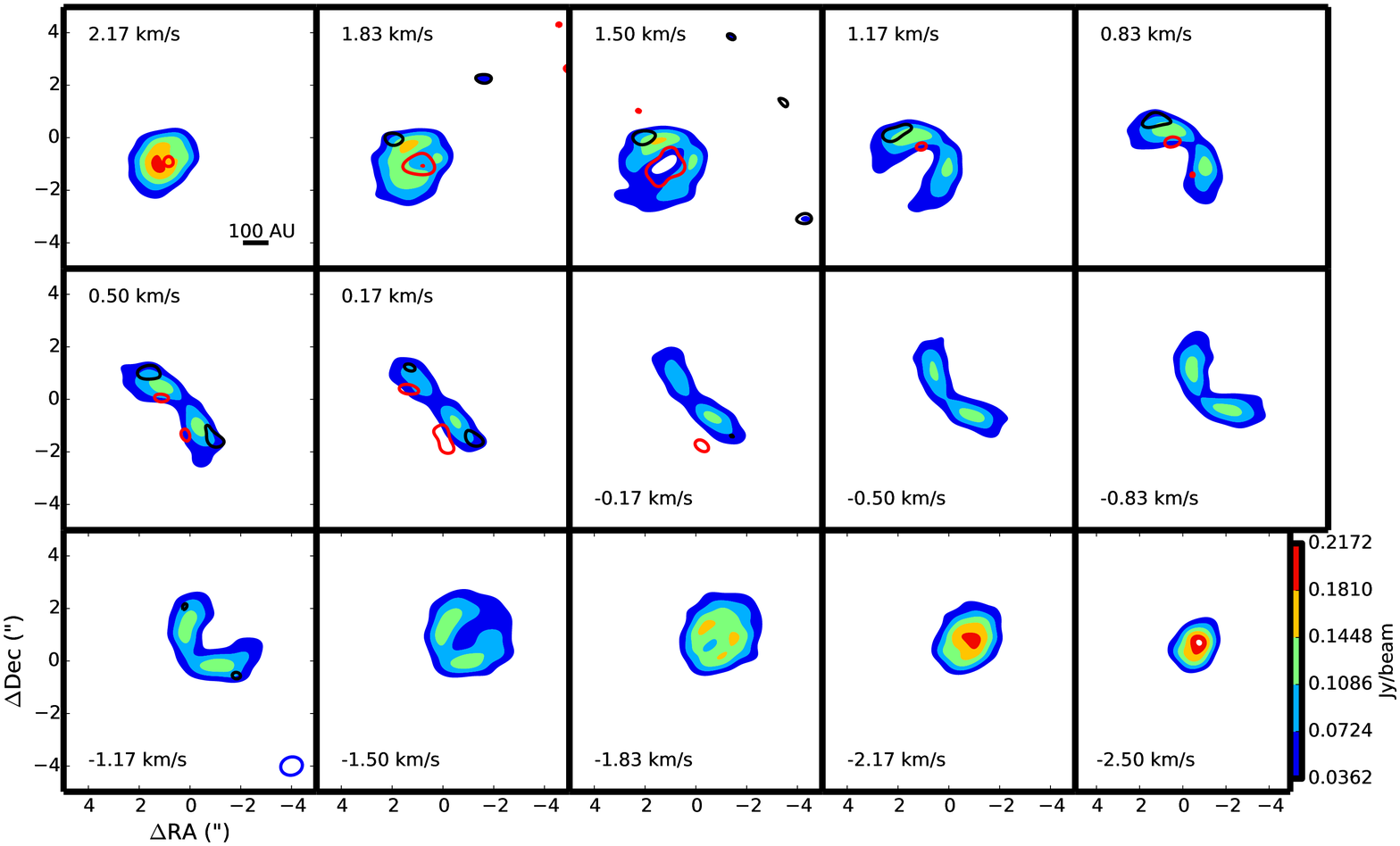}
\caption{Channel maps for the model (top, solid black contour) and the image generated from the difference in visibilities (bottom,red are model$>$data, black are data$>$model) for the fit to the C$^{18}$O(2-1) data (filled contours). Contours are in units of 0.036Jy/beam (=5$\sigma$). This model is able to successfully reproduce much of the emission.\label{chmaps_c18o21_vcs}}
\end{figure}

C$^{18}$O, being more heavily depleted than $^{13}$CO, is fully optically thin and is able to trace the entire vertical extent of the disk. As with $^{13}$CO, we break the degeneracy between density and temperature by using the derived values of $q$, $T_{\rm mid0}$, and $T_{\rm atm0}$ from CO(3-2), while allowing the depletion and $\gamma$ to vary. The resulting PDFs are shown in Figure~\ref{pdfs_c18o21} with channel maps and residuals in Figure~\ref{chmaps_c18o21_vcs}. We find a significant degeneracy between the depletion factor and $\gamma$ which is not surprising since both control the surface density of the disk. This anti-correlation may contribute to the difference in $\gamma$ between the fit to C$^{18}$O and the fit to $^{13}$CO; C$^{18}$O implies a steeper surface density profile, but this can be brought in line with the $^{13}$CO result with a depletion factor of $\sim$1700. We note that the C$^{18}$O line is fit with a higher inclination than the other lines. This difference is small, but significant. Further work is needed to fully explore the nature of this discrepancy. 

\begin{figure}
\center
\includegraphics[scale=.3]{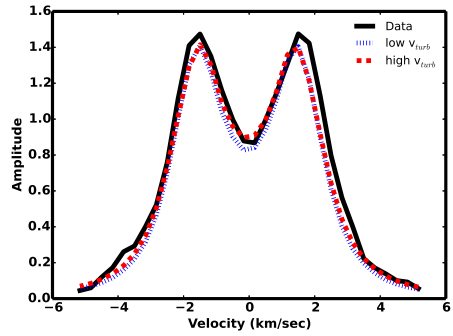}
\caption{Visibility spectra comparing the data (black solid line) to the best fit model with turbulence fixed at a low value (v$_{\rm turb}$=0.038c$_s$, blue dotted lines) and the best fit model with high turbulence (v$_{\rm turb}$=0.4c$_s$, red dashed line). The difference in spectra is small compared to the large difference in turbulence. \label{hiloturb_c18o21}}
\end{figure}

The C$^{18}$O line, with the lowest signal to noise, places the weakest constraint on the turbulent broadening, $v_{\rm turb}<0.4$c$_s$. At the coldest layers just above the CO freeze-out temperature, this implies random velocity dispersions of less than 150m s$^{-1}$. To check if a low turbulence model can adequately fit the data we ran an additional MCMC trial with the turbulence fixed at the low-resolution CO(3-2) limit ($v_{\rm turb}$=0.04c$_s$) and the results are shown in Figure~\ref{hiloturb_c18o21}. The difference between the high turbulence fit and the low turbulence fit is small despite the order of magnitude difference in turbulence. While C$^{18}$O is more sensitive to the midplane, its optically thin nature makes it less sensitive to velocity broadening from turbulence. In an optically thick line, the velocity shear provided by turbulence allows more photons to escape, increasing the flux, while in optically thin lines this effect is much smaller \citep{hor86}. This is consistent with the behavior seen in Figure~\ref{hiloturb_c18o21} for C$^{18}$O, as well as the weak limit on turbulence in $^{13}$CO.

\subsection{Multi-line fit}
\begin{figure}
\includegraphics[scale=1.]{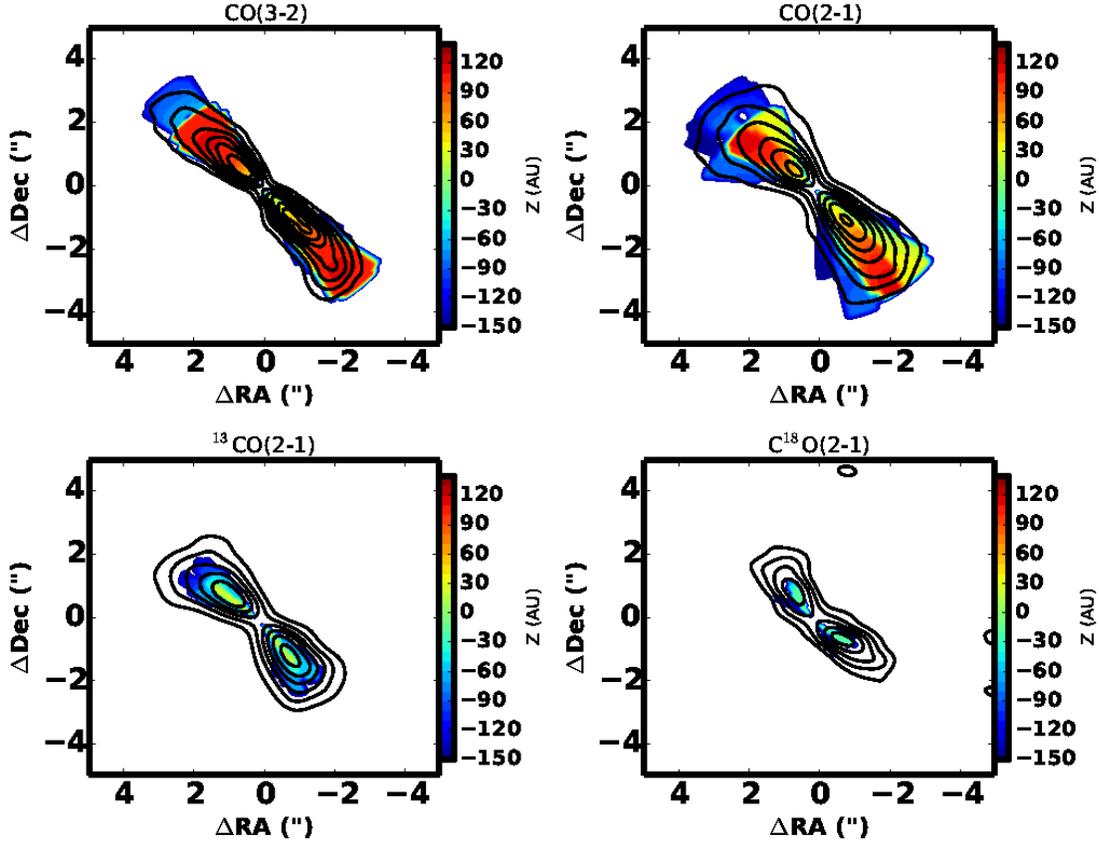}
\caption{Height of the $\tau=1$ surface for the best fit model derived for each line (filled contours). Overplotted are the observations for each line (similar contour levels as in Figures~\ref{chmaps_co32},\ref{chmaps_co21},\ref{chmaps_13co21},\ref{chmaps_c18o21_vcs}). Regions where there is disk emission but no marked $\tau$=1 surface are optically thin. The different lines probe different heights within the disk, giving us a handle on turbulence and temperature as a function of height within the disk.\label{chmaps_tau1_all}}
\end{figure}

The four lines, with their different optical depths, probe different regions within the disk. Figure~\ref{chmaps_tau1_all} shows the height of the $\tau=1$ surface for the best fit models to each of the four lines. At large radii, CO(3-2) is sensitive to materal at $z\sim$100AU, while $^{13}$CO and C$^{18}$O probe material closer to the midplane. By fitting one model to all four lines we can leverage this complementary information to more accurately constrain the temperature, density, and turbulence throughout the disk. For this MCMC trial we employ 60 walkers over 1000 steps, with the first 500 removed as burn-in. We fix $M_{\rm gas}$=0.09M$_{\odot}$ but allow $q$, $R_c$, $v_{\rm turb}$, $\gamma$, $T_{\rm atm0}$, $T_{\rm mid0}$, and inclination to vary, along with the CO/H$_2$ abundance. 
We find that allowing the CO/H$_2$ to vary, instead of the individual $^{13}$CO and C$^{18}$O depletion factors or the gas mass, provides the best multiline fit and we report those results here. Numerical simulations of MRI predict strong vertical gradients in the turbulent motion as a function of height within the disk \citep{fro06,sim13} but given the lack of detection found with the other lines we do not attempt to include any change in turbulence with height, beyond allowing it to vary with the local sound speed.

\begin{figure}
\includegraphics[scale=.5]{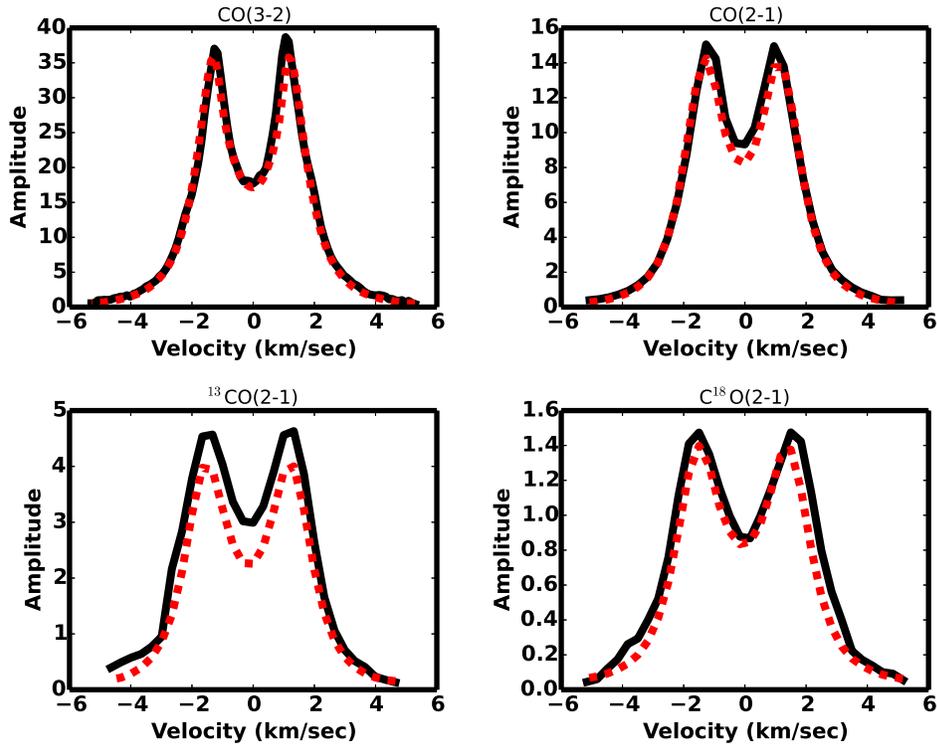}
\caption{Visibility spectra for all four emission lines (dark) along with the median model (red-dashed). The model fits have weak turbulence (v$_{\rm turb}<$0.16c$_s$) and are able to consistently fit all four lines, although it does underestimate $^{13}$CO.\label{spec_all}}
\end{figure}

\begin{figure}
\includegraphics[scale=.5]{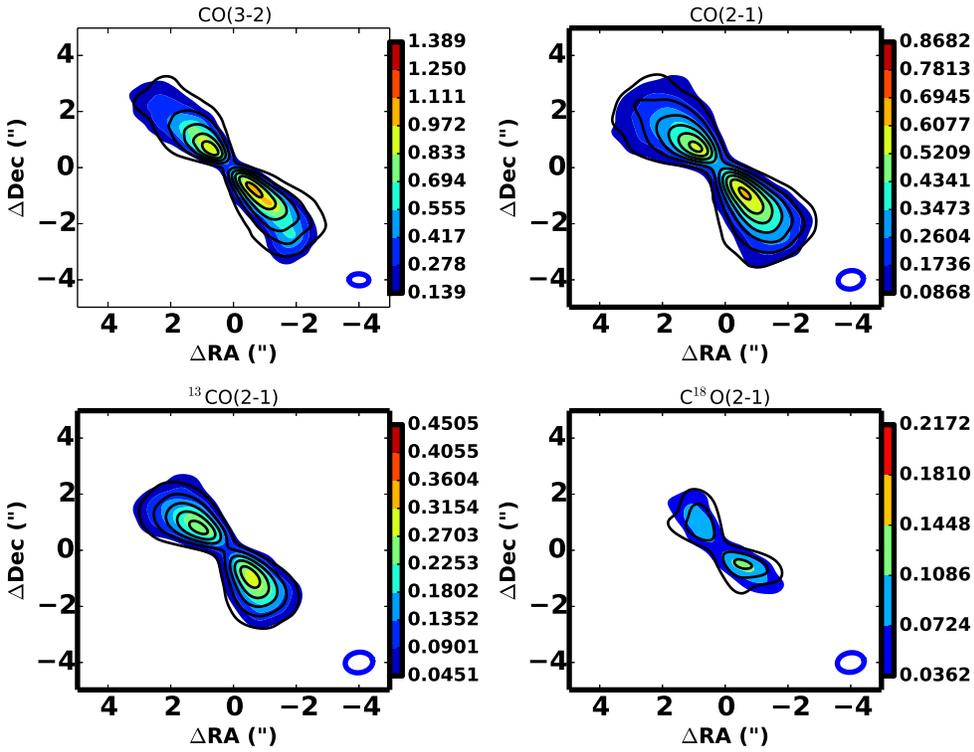}
\caption{Emission from the central velocity channel of all four lines (colored filled contours) compared to the model defined by the median of the PDFs (black lines). The model images are well-matched to the data, demonstrating that a model with low-turbulence is able to accurately fit all four lines. \label{chmaps_all}}
\end{figure}

\begin{figure}
\includegraphics[scale=.5]{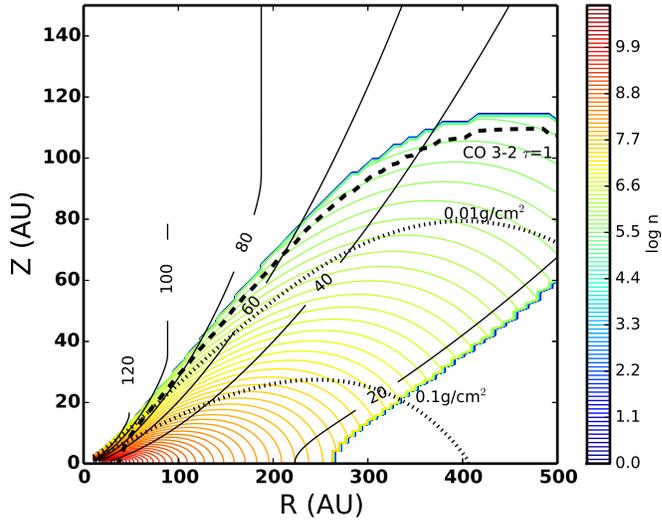}
\caption{Gas density (colored contours) and temperature (black contours) for the best model fit to all four CO lines. The density is only marked in the region of the disk where CO exists. \label{disk_struct_all}}
\end{figure}

The results are listed in Table~\ref{fit_all} with visibility spectra shown in Figure~\ref{spec_all}, and maps of the central velocity channel for each line in Figure~\ref{chmaps_all}. In terms of the overall density and temperature structure, while some of the parameters change compared to where they stood with the individual line fits, the overall structure is similar. Figure~\ref{disk_struct_all} shows the density and temperature structure for the model defined by the median of the PDFs, which is similar to the high-resolution CO(3-2) line result. We find that CO is depleted by a factor of 4.9$^{+0.6}_{-0.4}$ relative to the canonical ISM value. The individual fits to $^{13}$CO and C$^{18}$O suggest a depletion of only a factor of $\sim$3, which is smaller than that derived here because of the lower midplane temperature assumed in those fits. Recent observations have found mixed results when measuring CO/H$_2$ \citep[e.g.][]{fav13,fra14}, while models suggest that complex gas-grain chemistry in the cold midplane can affect the CO/H$_2$ ratio \citep{fur14,reb15}. It is also possible that the deviation in CO/H$_2$ is a sign that the assumed disk mass is incorrect. The mass was derived from the sub-mm emission of the dust \citep{ise07} and could deviate from the true value if there is a change in the gas to dust mass ratio, a concentration of dust into very large grains that are invisible in the sub-mm, or if the dust does not follow the gas distribution. We have run additional trials fixing the CO/H$_2$ ratio while allowing the disk mass to vary and find $M_{\rm gas}$=0.08$\pm$0.01M$_{\odot}$, which is similar to our assumed value, but is subject to large degeneracies, especially with $R_c$ and produces a worse overall fit to the data. Further work, including a more physically realistic treatment of CO chemistry near the ice line, is needed to confirm and understand the low CO/H$_2$ ratio.

The multi-line fit demonstrates that one consistent model, with weak turbulence, can fit all four lines. We find a three-sigma upper limit of $<$0.16c$_s$, which corresponds to non-thermal motion of $\sim$70-130m s$^{-1}$ in the upper layers and $\sim$50 m s$^{-1}$ at the CO freeze-out temperature. This again falls below the spectral resolution of the data, but the high spatial resolution and high signal to noise help to constrain turbulence to such low values. The weaker constraint on turbulence may be due to the isotopologues, with their lower spatial resolution and increased tolerance for high turbulence models, countering the pull of CO(3-2) towards low turbulence.

When fitting the CO(3-2) line by itself, we kept the CO/H$_2$ abundance fixed at the ISM value of 10$^{-4}$, while our multiline fit finds that the abundance is best fit with a value a factor of 4.9$^{+0.6}_{-0.4}$ lower. To see if this changes our initial result, we rerun the low-resolution CO(3-2) fit using the new abundance derived from the multiline fit. We find $v_{\rm turb}<$0.02c$_s$, consistent with the single line fit. 

In fitting just the CO(2-1) line we found that the turbulence measurement could be inflated due to unaccounted for complexities in the temperature structure. A similar effect can be seen here when including parameters for the systematic uncertainty, or when varying disk mass instead of CO abundance. We account for the systematic uncertainty by introducing two additional free parameters, one for the band 6 lines (CO(2-1), $^{13}$CO(2-1), C$^{18}$O(2-1)) and one for the band 7 line (CO(3-2)) since the two bands were calibrated separately. Each parameter is subject to a Gaussian prior with a dispersion of 20\%. Accounting for the systematic uncertainty results in a better overall fit to the data ($\chi^2_{\nu}$=0.9186 vs 0.9192 when s=1, ammounting to a $>10\sigma$ difference) and a stronger constraint on the turbulence ($v_{\rm turb}<$0.13c$_s$ vs $<$0.16c$_s$ when s=1). Allowing the disk mass to vary, instead of the CO/H$_2$ ratio, and including systematic uncertainty ends up with a model that is a worse overall fit ($\chi^2_{\nu}$= 0.9210) and has a weaker constraint on turbulence ($v_{\rm turb}<$0.21c$_s$). Going further and removing the systematic uncertainty parameters from the variable disk mass model results in an even worse fit ($\chi^2_{\nu}$= 0.9226) and an even softer limit on turbulence ($v_{\rm turb}<$0.24c$_s$). This progression suggests that as the models get worse at representing the data turbulence gets inflated in an effort to account for the deficiencies in the underlying structure. Such an effect can even be seen in the CO(3-2) low-resolution single line fit, where reducing the CO/H$_2$ abundance produces a better fit compared to the fiducial model (at $>$10$\sigma$ significance) with a stronger limit on the turbulence ($<$0.02c$_s$ vs $<$0.04c$_s$ for the fiducial fit). While our final best-fit model is certainly not a perfect representation of the HD 163296 system, a better model would likely lead to even more stringent constraints on turbulence.

\section{Implications of weak turbulence\label{turbulence}}
The high(low)-resolution CO(3-2) line puts an upper limit of only 3.1\%c$_s$ (3.8\%c$_s$) on the turbulence velocity dispersion in the outer disk, with a small increase to $\sim$5-6\%\ when accounting for uncertainties in distance and flux calibration. The CO(2-1), $^{13}$CO(2-1), and C$^{18}$O(2-1) lines offer less stringent upper limits ($<31\%$c$_s$,$<34\%$c$_s$,$<40\%$c$_s$ respectively) due to their lower spatial resolution. Simple tests fitting CO(2-1) and C$^{18}$O(2-1) with turbulence fixed at 3.8\%c$_s$ show that both lines can be well fit with low turbulence models, which we also found in a combined multiline fit that limited turbulence to $<$16\%\ c$_s$. In the case of optically thick lines like CO(2-1) subtleties in the temperature structure can be masked by turbulence and in the case of optically thin lines like C$^{18}$O(2-1) the profile is not particularly sensitive to turbulence. The upper limit from CO(3-2) corresponds to $\sim$9-19m s$^{-1}$ across the upper layers of the CO emitting region while the more optically thin $^{13}$CO and C$^{18}$O limit non-thermal motion close to the CO freeze-out zone at the midplane to $<$150m s$^{-1}$, although the multiline fit pushes that limit down to $<$50m s$^{-1}$. While this falls below the spectral resolution of the data, fitting the full three-dimensional data set includes the spatial information that drives the limit to such low values. 

The viscosity associated with turbulence is often parameterized by $\alpha$ which encapsulates much of the detailed physics relating the largest velocity and spatial scales, c$_s$, and H, to the turbulent viscosity ($\nu=\alpha c_s H$) \citep{sha73}. In a turbulent cascade, energy is deposited on the largest spatial scales and cascades down to smaller scales \citep{kol41}. Even though at the spatial resolution of our CO(3-2) data ($\sim$60AU) we cannot resolve the largest possible spatial scale (H$\sim$5AU at 100AU), we can see the influence of turbulence on the velocity structure. Here $\alpha$ entails the strength and size of the cascade relative to H and c$_s$ which represent the largest spatial and velocities scales for the turbulence. The parameter $\alpha$ can also be related directly to the stress tensor W$_{R\phi}$, which is the density weighted sum of the Maxwell and Reynolds stresses \citep{bal98}. This tensor encapsulates the correlated fluctuations in the R-$\phi$ plane and is connected to the viscous evolution through $\alpha$=W$_{R\phi}$/c$_s^2$. The form of this equation suggests that $\alpha\sim$($v_{\rm turb}$/c$_s$)$^2$, consistent with simulations that measure both turbulent motion and the stresses \citep[e.g.][]{sim13,sim15}. With this conversion from turbulent velocity to $\alpha$, our CO(3-2) observations correspond to an upper limit of $\alpha<9.6\times10^{-4}$ while the multiline fit provides a limit of $\alpha<0.03$. 

Our limits indicate that turbulence is weaker than previous measurements of turbulent motion in the disk around HD 163296. \citet{hug11} report a tentative 3$\sigma$ detection of 300m s$^{-1}$ turbulence for the HD 163296 system based on SMA data with higher spectral, but lower spatial, resolution, while \citet{deg13} find that a model with 0.1km s$^{-1}$ turbulence fits the ALMA dataset better than a model with no turbulence. Our CO(3-2) limit corresponds to $\sim$9-19m s$^{-1}$ across the upper layers of the CO emitting region. While the best fit model in \citet{hug11} is able to fit the SMA data, the limited constraint on the temperature structure means that this data is subject to a stronger degeneracy between temperature and turbulence. In particular, the low spatial resolution SMA data has difficulty distinguishing between a cold, flat disk with large turbulent broadening and a warm, puffy disk with small turbulent broadening. By separating the front and back side of the disk we can limit the models to warmer disks with weaker turbulence.

The difference between our upper limit and \citet{deg13} likely is a result of different disk models. They employ a radiative transfer code to calculate the dust temperature and then use this to calculate the gas emission subject to assumptions about equal gas and dust temperatures, constant gas to dust mass ratio, the grain size distribution and the abundance of CO relative to H$_2$, subject to freeze-out and photodissociation. Our disk models focus solely on the gas structure making them simpler although at the expense of the additional information about disk structure that comes from the dust emission. The dust will mainly influence the gas temperature; the dust dominates the opacity and absorbs thermal energy from stellar radiation transferring it through collisions to the gas. By assuming a functional form for the temperature structure we avoid the complicated radiative transfer problem relating the dust and stellar emission. Efforts to use the dust to constrain the gas structure have been complicated by recent observations that have found evidence of different gas and dust distributions in protoplanetary disks. \citet{and12} find that the dust disk within TW Hya is much more compact than the gas, even when accounting for differences in optical depth. \citet{deg13} note that their fit to the gas and dust in the HD 163296 system requires different surface density gradients. This suggests that a more detailed treatment is needed to fully reconcile the dust and gas emission. The complexity of the models used in \citet{deg13} do not allow for a full characterization of the uncertainties on the various parameters, which allows for the possibility that our results are consistent with theirs to within the uncertainties. 

Our limits also fall below measurements in other stars. \citet{har98} find $\alpha\sim0.01$ based on the time evolution of accretion and typical disk masses. Near-infrared CO overtone emission, sensitive to the innermost regions of circumstellar disks, has found transonic line broadening consistent with strong turbulence \citep{naj96,car04,naj09}, although without the benefit of spatial information these observations are subject to stronger degeneracies. It may be that the turbulence within the inner regions of the disk probed by the overtone emission and the accretion rate onto the star is much higher than in the outer disk. Close to the star thermal ionization of akali metals is expected to be strong enough to activate MRI without strong damping from non-ideal MHD effects, such as ambipolar diffusion, that are expected to play a larger role in the outer disk \citep{arm11}. Recent observations suggest that the cosmic ray ionization rate, important in the outer disk, may be lower than previously expected \citep{cle13}, further damping the influence of MRI in the outer disk. Electric fields created by MRI may accelerate charged particles, enhancing recombination collisions with dust grains, reducing the ionization and in turn damping MRI in the outer disk \citep{mor15}. Our results are consistent with the limit on turbulent velocity found in TW Hya \citep{hug11}. \citet{gui12} measure the turbulence in DM Tau to be 130m s$^{-1}$, which, at the typical height of the CS emission (z/r$\sim$0.1-0.2), is similar to our limits from $^{13}$CO and C$^{18}$O ($\lesssim$150m s$^{-1}$) probing a similar area within the disk, although CS may be subject to the same optically thin sensitivity issues as C$^{18}$O. 

Numerical simulations of the MRI, with background fields optimized for full-blown MRI turbulence in far-ultraviolet ionized layers, find that turbulent motions vary from a few percent of the sound speed near the midplane up to values comparable to the sound speed at 3-5 scale heights above the midplane, at surface densities $\lesssim$ 0.1 g cm$^{-2}$ \citep{sim13,fro06,per11,sim15}. The CO(3-2) $\tau=1$ surface in our model disks lies at z=30-100AU (Figure~\ref{chmaps_tau1_all}), which corresponds to 3H far from the star and 5H close to the star, or equivalently surface densities $\sim$5$\times$10$^{-3}$ g cm$^{-2}$. \citet{sim15}, in their numerical simulations of a disk modeled after the HD 163296 system, find that at these heights/surface densities at 100 AU, far-ultraviolet (FUV) photons are able to effectively ionize the disk leading to strong turbulence. The distribution of velocities peaks at $v_{\rm turb}$/c$_s$$\sim$0.4-1, depending on the exact physical conditions and the exact height within the disk. The full range of velocities follows a Maxwell-Boltzmann distribution, with velocities ranging over a factor of $\sim$4 about the peak. Our CO(3-2) upper limit falls well below this expected level of turbulence high in the disk. Close to the midplane the weaker ionization, due to the attenuation of FUV photons, and higher densities, lead to lower Alfv\'{e}n speeds, which in turn is associated with weaker turbulence, with typical velocities of $v_{\rm turb}$/c$_s\sim$0.03-0.06 \citep{sim15}. The $\tau=1$ C$^{18}$O surface pierces through the disk to reach z=H on the bottom half in the disk, making it more sensitive to the midplane than CO(3-2). The less stringent limit provided by this line is consistent with the expected weak turbulence near the midplane, although as discussed earlier optically thin lines have a weaker response to velocity shear, limiting the ability of C$^{18}$O to constrain turbulence. Simulations have also found that a net vertical magnetic field is required to produce strong turbulence \citep{bai11,sim13}; in the HD 163296 system the MRI may still operate but simply with a weak magnetic field that results in weak turbulence. An MRI `dead zone' due to Ohmic diffusion may be present near the midplane where the ionization is insufficient to effectively couple the gas to the magnetic field, although these zones are expected to extend out to $R\sim10-30$AU \citep[e.g.][]{san00,dzy13}, while our measurements are most sensitive to the behavior in the outer ($R>30$AU) disk.

Turbulence is connected with the accretion rate through angular momentum exchange, with higher levels of turbulence corresponding to more vigorous accretion. In this context the weak turbulence is surprising given the moderately high accretion rate onto HD 163296 of 5$\times10^{-7}$M$_{\odot}$ yr$^{-1}$ \citep{men13}. Assuming this accretion rate applies throughout the disk, and that there are no external torques, $\alpha$ is directly related to $\dot{M}$ and the temperature and density structure of the disk \citep{arm11}.  This accretion rate, combined with a disk mass of 0.09M$_{\odot}$ and a critical radius of 175AU, implies $\alpha$ increasing from 0.06 at 10 AU to 0.3 at 500 AU while our observations suggest $\alpha$ is less than 9.6$\times$10$^{-4}$. The low disk mass and low turbulence appear to be incompatible with the high accretion rate onto the star. An uncertainty in the disk mass will factor into this estimate of $\alpha$, but the disk mass would need to be $\sim$10M$_{\odot}$ to bring $\alpha$ in line with our measurements. One possible explanation is that the viscosity increases strongly toward the inner disk, leading to a higher accretion rate onto the star than in the outer disk. At the low densities in the outer disk ambipolar diffusion dominates while closer to the star ohmic dissipation and the hall effect play a larger role \citep{arm11,tur14}. It may also be the case that the disk is not in steady state equilibrium. \citet{men13} find that the current accretion rate is at least an order of magnitude higher than in observations taken 15 years earlier. Given a disk mass of 0.09M$_{\odot}$, an accretion rate of 5$\times10^{-9}$M$_{\odot}$ yr$^{-1}$, two orders of magnitude lower than the recent observations, could reproduce the observed $\alpha$ in the outer disk.

Another explanation for the high accretion rate with low turbulence is if there is a form of angular momentum removal that does not lead to an observable velocity dispersion, such as a disk wind \citep[e.g.][]{bla82}. Numerical models find that magneto-hydrodynamics winds can be driven from the disk, although the exact results depend strongly on the numerical setup \citep{fro13}. \citet{kla13} observe evidence for a disk wind from HD 163296 based on the morphology of large scale CO emission. Further observations are needed to determine if this wind originates in the outer disk, and if it removes enough angular momentum to explain the accretion rate.

Gravito-turbulence has been suggested as an alternative to MRI as the driver of turbulence within the disk \citep{arm11}, although as with MRI the predicted magnitude is still larger than our upper limits. \citet{shi14} predict $v_{\rm turb}$/c$_s$$\sim$0.4 throughout the disk while SPH simulations by \citet{for12} predict velocity distributions that peak at v$_{\rm turb}$/c$_s$$\sim$0.1 in the plane of the disk, with motions an order of magnitude slower in the vertical direction. While anisotropic turbulence could diminish the measured motions, the inclination of the HD 163296 disk means that motions in the plane of the disk still contribute substantially to the observed velocity dispersion, limiting the influence of any weak motion in the vertical direction. The large velocities predicted by gravito-turbulence are ruled out by our CO(3-2) observations (gravito-turbulence is not in any case favored because the disk mass is nominally small).

Given the very weak turbulence, it is possible that hydrodynamic instabilities start to play a role. Vortices created by the baroclinic instability have been proposed as a mechanism for transporting angular momentum through the disk \citep{kla03}. In regions of weak MRI, a vertical shear instability may develop in the presence of varying angular velocity with height resulting in Reynolds stresses that drive $\alpha\sim10^{-3}$ \citep{nel13}. Other hydrodynamic instabilities \citep[e.g.][]{lyr14,mar14} may also play a role when MRI is weak, depending on the exact structure of the disk. Further modeling is needed to determine if the physical conditions necessary for these instabilities are present in this disk. 

Such low turbulence can lead to measurable effects on the dust distribution and the chemistry in the disk. The dust vertical density distribution in a turbulent disk is particularly sensitive to the gas velocities in the upper layers of the atmosphere \citep{fro09} with diminished velocities leading to increased settling \citep{dub95,joh05,car06,cie07}. Our CO(3-2) measurements find weak turbulent motion in the upper atmosphere, suggesting that dust settling should be significant and indeed HD 163296 has been observed to have a diminished mid-infrared flux consistent with substantial settling of small dust grains \citep{mee01,juh10}. The chemical structure may contain imprints of turbulence since large turbulent velocities lead to enhanced mixing and a smoothing of chemical gradients. This can be especially important near sharp boundaries, such as ice lines. Given that turbulence operates over a typical length scale of H, it more strongly affects the steeper vertical chemical gradients than the shallower radial gradients \citep{sem11} and is unlikely to strongly affect the radial location of e.g. the CO snow line. \citet{fur14} find that models with weak turbulence can lead to substantial depletion of CO, since a more laminar structure allows for the slow sink of carbon out of CO and into carbon-chain molecules. This is consistent with our evidence of diminished CO/H$_2$ although more detailed modeling is needed to distinguish this complex chemistry from other effects such as selective photodissociation.

\section{Conclusions}
Using ALMA science verification data of CO(3-2), CO(2-1), $^{13}$CO(2-1), and C$^{18}$O(2-1) we constrain the turbulent velocity dispersion within the protoplanetary disk surrounding the young star HD 163296. By employing physically realistic models and a Bayesian analysis we find a limit from the high-resolution CO(3-2) data of $v_{\rm turb}<$0.031c$_s$, which implies $\alpha<$9.6$\times$10$^{-4}$, with consistent results when fitting the other lines individually, or together with the CO(3-2) data. This tight constraint comes from the high spatial resolution, which can both probe the temperature structure and directly constrain the turbulence in a way that was not possible with previous instruments. Our upper limit is well below that expected from the strong accretion onto the star itself, suggesting either that angular momentum transport is strongly radially dependent or that angular momentum is removed in a way that does not generate significant non-thermal motion. Our limits also fall below predictions from numerical simulations of the MRI, suggesting that MRI is less efficient in the outer disk than previously thought. This demonstrates that with the high spatial resolution and sensitivity of ALMA we can place physically meaningful constraints on the turbulence. 

\acknowledgements
We thank the referee for comments that helped improve the quality of the paper. We gratefully acknowledge support from NASA Origins of Solar Systems Grant NNX13AI32G. We also thank Chunhua Qi and Karin Oberg for helpful discussions regarding the structure of CO. This paper makes use of the following ALMA data: ADS/JAO.ALMA\#2011.0.00010.SV. ALMA is a partnership of ESO (representing its member states), NSF (USA) and NINS (Japan), together with NRC (Canada), NSC and ASIAA (Taiwan), and KASI (Republic of Korea), in cooperation with the Republic of Chile. The Joint ALMA Observatory is operated by ESO, AUI/NRAO and NAOJ. The National Radio Astronomy Observatory is a facility of the National Science Foundation operated under cooperate agreement by Associate Universities, Inc. This research made use of Astropy, a community-developed core Python package for Astronomy \citep{ast13}.

\begin{deluxetable}{ccccccccccc}
\tablewidth{0pc}
\rotate
\tablecaption{Model Fitting Results\label{standard}}
\tablehead{ \colhead{Line} & \colhead{$q$} & \colhead{$R_c$(AU)} & \colhead{$v_{\rm turb}$/c$_{\rm s}$}  &  \colhead{$T_{\rm atm0}$(K)} & \colhead{$\gamma$} & \colhead{$T_{\rm mid0}$(K)} & \colhead{inclination} & \colhead{N(X)/N($^{12}$C$^{16}$O)} & \colhead{ $\chi_{\nu}^2$}\tablenotemark{a}}
\startdata
CO(3-2) (high-res) & -0.278$^{+0.003}_{-0.008}$ & 175$^{+13}_{-4}$ & $<$0.031 &  86$\pm1$ & 1\tablenotemark{b} & 21.8$^{+0.7}_{-0.4}$ & 47.6$\pm0.1$ & 1 & 0.96\\
CO(3-2) (low-res) & -0.216$^{+0.01}_{-0.008}$ & 194$^{+12}_{-15}$ & $<$0.038 &  94$\pm2$ & 1\tablenotemark{b} & 17.5$^{+0.8}_{-0.7}$ & 48.3$\pm0.3$ & 1 & 0.94\\
CO(2-1) & -0.27$^{+0.02}_{-0.01}$ & 224$^{+21}_{-16}$ & $<$0.31 & 79.0$^{+1.1}_{-1.5}$ & 1\tablenotemark{b} & 17.5\tablenotemark{b} & 46.1$^{+0.4}_{-0.3}$  & 1 & 0.92\\
$^{13}$CO(2-1) & -0.216\tablenotemark{b} & 194\tablenotemark{b} & $<$0.34 & 94\tablenotemark{b} & 0.78$^{+0.02}_{-0.01}$ & 17.5\tablenotemark{b} & 47.5$\pm0.2$ & 233$^{+3}_{-15}$ & 0.94\\
C$^{18}$O(2-1) & -0.216\tablenotemark{b} & 194\tablenotemark{b} & $<$0.40 & 94\tablenotemark{b} & 1.13$\pm0.08$ & 17.5\tablenotemark{b} & 51.5$^{+0.2}_{-0.3}$ & 1240$^{+120}_{-110}$ & 0.92\\
\enddata
\tablecomments{Median plus 3$\sigma$ ranges for the PDFs defined by the walker positions after removing burn-in.}
\tablenotetext{a}{Reduced chi-squared for the model defined by the median of the PDFs}
\tablenotetext{b}{Held fixed during the model fitting.}
\end{deluxetable}

\begin{deluxetable}{cc}
\tablewidth{0pc}
\tablecaption{Multi-line model fit\label{fit_all}}
\tablehead{ \colhead{Parameter} & \colhead{Result}}
\startdata
$q$ & -.298$^{+0.017}_{-0.008}$\\
$R_c$ & 195$^{+2}_{-17}$\\
$v_{\rm turb}$ & $<0.16$c$_s$\\
$T_{\rm atm0}$ & 85.5$^{+1.2}_{-0.8}$\\
$\gamma$ & 0.89$^{+0.07}_{-0.02}$\\
$T_{\rm mid0}$ & 22.5$^{+0.5}_{-0.6}$\\
incl & 48.4$^{+0.1}_{-0.4}$\\
log(CO/H$_2$) & -4.69$^{+0.05}_{-0.04}$\\
$\chi^2_{\nu}$ & 0.92\tablenotemark{a}\\
\enddata
\tablecomments{Median plus 3$\sigma$ ranges for the PDFs defined by the walker positions after removing burn-in.}
\tablenotetext{a}{Reduced chi-squared from the combination of all four lines for the model defined by the median of the PDFs.}
\end{deluxetable}
\clearpage

\clearpage

\clearpage

\clearpage

\clearpage

\begin{thebibliography}{}
\bibitem[Aikawa \&\ Herbst(1999)]{aik99} Aikawa, Y. \&\ Herbst, E. 1999, \aap, 351, 233
\bibitem[Andrews et al.(2010)]{and10} Andrews, S.M., Wilner, D.J., Hughes, A.M., Qi, C., \&\ Dullemond, C.P. 2010, \apj, 723, 1241
\bibitem[Andrews et al.(2012)]{and12} Andrews, S., et al. 2012, \apj, 744, 162
\bibitem[Armitage(2011)]{arm11} Armitage, P.J. 2011, \araa, 49, 195
\bibitem[Astropy Collaboration et 
al.(2013)]{ast13} Astropy Collaboration, Robitaille, T.~P., Tollerud, E.~J., et al.\ 2013, \aap, 558, A33 
\bibitem[Bai \&\ Stone(2011)]{bai11} Bai, X.-N., \& Stone, J.~M.\ 2011, \apj, 736, 144
\bibitem[Balbus \&\ Hawley(1991)]{bal91} Balbus, S.A. \&\ Hawley, J.F. 1991, \apj, 376, 214
\bibitem[Balbus \& Hawley(1998)]{bal98} Balbus, S.~A., \& Hawley, J.~F.\ 1998, Reviews of Modern Physics, 70, 1
\bibitem[Bergin et al.(1995)]{ber95} Bergin, E.A., Langer, W.D., \&\ Goldsmith, P.F. 1995, \apj, 441, 222
\bibitem[Blandford \& Payne(1982)]{bla82} Blandford, R.~D., \& Payne, D.~G.\ 1982, \mnras, 199, 883
\bibitem[Blum 
\& Wurm(2000)]{blu00} Blum, J., \& Wurm, G.\ 2000, \icarus, 143, 138
\bibitem[Blum 
\& Wurm(2008)]{blu08} Blum, J., \& Wurm, G.\ 2008, \araa, 46, 21
\bibitem[Carballido et al.(2006)]{car06} Carballido, A., 
Fromang, S., \& Papaloizou, J.\ 2006, \mnras, 373, 1633 
\bibitem[Carr et al.(2004)]{car04} Carr, J.S., Tokunaga, A.T., \&\ Najita, J. 2004, \apj, 603, 213
\bibitem[Ciesla(2007)]{cie07} Ciesla, F.J. 2007 \apj, 654, L159
\bibitem[Cleeves et al.(2013)]{cle13} Cleeves, L.~I., Adams, 
F.~C., \& Bergin, E.~A.\ 2013, \apj, 772, 5 
\bibitem[Crida et al.(2006)]{cri06} Crida, A., Morbidelli, 
A., \& Masset, F.\ 2006, \icarus, 181, 587
\bibitem[D'Alessio et al.(2006)]{dal06} D'Alessio, P., Calvet, N., Hartmann, L., Franco-Hern\'{a}ndez, R., \&\ Serv\'{i}n, H. \apj, 638, 314
\bibitem[Dartois et al.(2003)]{dar03} Dartois, E., Dutrey, A., \&\ Guilloteau, S. 2003, \aap, 399, 773
\bibitem[de Gregorio-Monsalvo et al.(2013)]{deg13} de Gregorio-Monsalvo, I., et al. 2013, \aap,, 557, 133
\bibitem[Dubrulle et al.(1995)]{dub95} Dubrulle, B., Morfill, 
G., \& Sterzik, M.\ 1995, \icarus, 114, 237
\bibitem[Dzyurkevich et al.(2013)]{dzy13} Dzyurkevich, N., 
Turner, N.~J., Henning, T., \& Kley, W.\ 2013, \apj, 765, 114
\bibitem[Favre et al.(2013)]{fav13} Favre, C., Cleeves, L.I., Bergin, E.A., Qi, C., \&\ Blake, G.A. 2013, \apj, 776L, 38
\bibitem[Flock et al.(2011)]{flo11} Flock, M., Dzyurkevich, 
N., Klahr, H., Turner, N.~J., \& Henning, T.\ 2011, \apj, 735, 122
\bibitem[Foreman-Mackey et al.(2013)]{for13} Foreman-Mackey, D., Hogg, D.W., Lang, D., \&\ Goodman, J. 2013, \pasp, 125, 306
\bibitem[Forgan et al.(2012)]{for12} Forgan, D., Armitage, P.~J., \& Simon, J.~B.\ 2012, \mnras, 426, 2419
\bibitem[France et al.(2014)]{fra14} France, K., Herczeg, G.J., McJunkin, M., \&\ Pention, S.V. 2014, \apj, 794, 160
\bibitem[Fromang \&\ Nelson(2006)]{fro06} Fromang, S. \&\ Nelson, R.P. 2006, \aap, 457, 343
\bibitem[Fromang 
\& Nelson(2009)]{fro09} Fromang, S., \& Nelson, R.~P.\ 2009, \aap, 496, 597 
\bibitem[Fromang et al.(2013)]{fro13} Fromang, S., Latter, H., Lesur, G., \&\ Ogilvie, G.I. \aap, 552, 71
\bibitem[Fung et al.(2014)]{fun14} Fung, J., Shi, J.-M., 
\& Chiang, E.\ 2014, \apj, 782, 88
\bibitem[Furuya 
\& Aikawa(2014)]{fur14} Furuya, K., \& Aikawa, Y.\ 2014, \apj, 790, 97
\bibitem[Gammie(1996)]{gam96} Gammie, C.F. 1996, \apj, 457, 355
\bibitem[Goodman \&\ Weare(2010)]{goo10} Goodman, J., \&\ Weare, J. 2010, Commun. Appl. Math. Comput. Sci., 5, 65
\bibitem[Gressel et al.(2011)]{gre11} Gressel, O., Nelson, 
R.~P., \& Turner, N.~J.\ 2011, \mnras, 415, 3291
\bibitem[Guilloteau et al.(2012)]{gui12} Guilloteau, S., Dutrey, A., Wakelam, V., Hersant, F., Semenov, D., Chapillon, E., Henning, T., \&\ Pi\'{e}tu, V. 2012, \aap, 548, 70
\bibitem[Gundlach et al.(2011)]{gun11} Gundlach, B., Kilias, S., Beitz, E., \&\ Blum, J. 2011, Icarus, 214, 717
\bibitem[Hartmann et al.(1998)]{har98} Hartmann, L., Calvet, N., Gullbring, E., \&\ D'Alessio, P. 1998, \apj, 495, 385
\bibitem[Hartmann et al.(2004)]{har04} Hartmann, L., Hinkle, K., \&\ Calvet, N. 2004, \apj, 609, 906
\bibitem[Hughes et al.(2008)]{hug08} Hughes, A.M., Wilner, D.J., Qi, C., \&\ Hogerheijde, M.R. 2008, \apj, 678, 1119
\bibitem[Hughes et al.(2011)]{hug11} Hughes, A.M., Wilner, D.J., Andrews, S.M., Qi, C., \&\ Hogerheijde, M.R. 2011, \apj, 727, 85
\bibitem[Ida et al.(2008)]{ida08} Ida, S., Guillot, T., \&\ Morbidelli, A. 2008, 686, 1292
\bibitem[Isella et al.(2007)]{ise07} Isella, A., Testi, L., Natta, A., Neri, R., Wilner, D., \&\ Qi, C. 2007, \aap, 469, 213
\bibitem[Johansen 
\& Klahr(2005)]{joh05} Johansen, A., \& Klahr, H.\ 2005, \apj, 634, 1353 
\bibitem[Johansen et al.(2007)]{joh07} Johansen, A., Oishi, J.~S., Mac Low, M.-M., et al.\ 2007, \nat, 448, 1022
\bibitem[Jonkheid et al.(2007)]{jon07} Jonkheid, B., Dullemond, C.P., Hogerheijde, M.R., \&\ van Dishoeck, E.F. 2007, \aap, 463, 203
\bibitem[Juh\'{a}sz et al.(2010)]{juh10} Juh\'{a}sz, A., et al. 2010, \apj, 721, 431
\bibitem[Klaassen et al.(2013)]{kla13} Klaassen, P.D., et al. 2013, \aap, 555, 73
\bibitem[Klahr 
\& Henning(1997)]{kla97} Klahr, H.~H., \& Henning, T.\ 1997, \icarus, 128, 213
\bibitem[Klahr \& Bodenheimer(2003)]{kla03} Klahr, H.~H., \& Bodenheimer, P.\ 2003, \apj, 582, 869
\bibitem[Kley 
\& Nelson(2012)]{kle12} Kley, W., \& Nelson, R.~P.\ 2012, \araa, 50, 211
\bibitem[Kolmogorov(1941)]{kol41} Kolmogorov, A.\ 1941, 
Akademiia Nauk SSSR Doklady, 30, 301
\bibitem[Horne \& Marsh(1986)]{hor86} Horne, K., \&\ Marsh, T.R. 1986, \mnras, 218, 761
\bibitem[Laughlin et al.(2004)]{lau04} Laughlin, G., Steinacker, A., \&\ Adams, F.C. 2004, 608, 489
\bibitem[Lynden-Bell \&\ Pringle(1974)]{lyn74} Lynden-Bell, D. \&\ Pringle, J.E. 1974, \mnras, 168, 603
\bibitem[Lyra et al.(2014)]{lyr14} Lyra, W., Turner, N., \& McNally, C.\ 2014, arXiv:1410.8092
\bibitem[Marcus et al.(2014)]{mar14} Marcus, P., Pei, S., Jiang, C.-H., et al.\ 2014, arXiv:1410.8143
\bibitem[Mathews et al.(2013)]{mat13} Mathews, G.S., et al. 2013, \aap, 557, 132
\bibitem[Meeus et 
al.(2001)]{mee01} Meeus, G., Waters, L.~B.~F.~M., Bouwman, J., et al.\ 2001, \aap, 365, 476
\bibitem[Mendigutia et al.(2013)]{men13} Mendigutia, I., et al. 2013, \apj, 776, 44
\bibitem[Miller \& Stone(2000)]{mil00} Miller, K.~A., \& Stone, J.~M.\ 2000, \apj, 534, 398
\bibitem[Miotello et al.(2014)]{mio14} Miotello, A., Bruderer, S., \& van Dishoeck, E.~F.\ 2014, \aap, 572, A96
\bibitem[Montesinos et al.(2009)]{mon09} Montesinos, B., Eiroa, C., Mora, A., \&\ Merin, B. 2009, \aap, 495, 901
\bibitem[Mori \&\ Okuzumi(2015)]{mor15} Mori, S. \&\ Okuzumi, S. 2015, arXiv:1505.04896
\bibitem[Najita et al.(1996)]{naj96} Najita, J.R., Carr, J.S., Glassgold, A.E., Shu, F.H., \&\ Tokunaga, A.T. 1996, \apj, 462, 919
\bibitem[Najita et al.(2009)]{naj09} Najita, J.R., Doppmann, G.W., Carr, J.S., Graham, J.R. \&\ Eisner, J.A. 2009, \apj, 691, 738
\bibitem[Nelson et al.(2013)]{nel13} Nelson, R.~P., Gressel, O., \& Umurhan, O.~M.\ 2013, \mnras, 435, 2610
\bibitem[Pavlyuchenkov et al.(2007)]{pav07} Pavlyuchenkov, Y., et al. 2007, \apj, 669, 1262
\bibitem[Perez-Becker \& Chiang(2011)]{per11} Perez-Becker, D., \& Chiang, E.\ 2011, \apj, 735, 8
\bibitem[Poppe et al.(2000)]{pop00} Poppe, T., Blum, J., \&\ Henning, T. 2000, 533, 454
\bibitem[Qi et al.(2011)]{qi11} Qi, C., et al. 2011, \apj, 740, 84
\bibitem[Qi et al.(2013a)]{qi13a} Qi, C., et al. 2013, Science, 341, 630
\bibitem[Qi et al.(2013b)]{qi13b} Qi, C., Oberg, K., Wilner, D.J., \&\ Rosenfeld, K.A. 2013, \apj, 765, 14
\bibitem[Reg{\'a}ly et 
al.(2011)]{reg11} Reg{\'a}ly, Z., S{\'a}ndor, Z., Dullemond, C.~P., \& Kiss, L.~L.\ 2011, \aap, 528, A93
\bibitem[Reboussin et al.(2015)]{reb15} Reboussin, L., Wakelam, V., Guilloteau, S., Hersant, F., \&\ Dutrey, A. 2015, arXiv:1505.01309
\bibitem[Rosenfeld et al.(2012)]{ros12} Rosenfeld, K.~A., Qi, C., Andrews, S.~M., et al.\ 2012, \apj, 757, 129
\bibitem[Rosenfeld et al.(2013)]{ros13} Rosenfeld, K.A., Andrews, S.M., Hughes, A.M., Wilner, D.J., \&\ Qi, C. 2013, \apj, 774, 16
\bibitem[Rosenfeld et al.(2013)]{ros13b} Rosenfeld, K.~A., Andrews, S.~M., Wilner, D.~J., Kastner, J.~H., \& McClure, M.~K.\ 2013, \apj, 775, 136
\bibitem[Rosenfeld et al.(2014)]{ros14} Rosenfeld, K.~A., Chiang, E., \& Andrews, S.~M.\ 2014, \apj, 782, 62
\bibitem[Sano et al.(2000)]{san00} Sano, T., Miyama, S.~M., 
Umebayashi, T., \& Nakano, T.\ 2000, \apj, 543, 486
\bibitem[Shakura \& Sunyaev(1973)]{sha73} Shakura, N.~I., \& Sunyaev, R.~A.\ 1973, \aap, 24, 337 
\bibitem[Shi \&\ Chiang(2014)]{shi14} Shi, J. \&\ Chiang, E. 2014, 78, 34
\bibitem[Simon et al.(2013)]{sim13} Simon, J.B., Bai, X., Armitage, P.J., Stone, J.M., \&\ Beckwith, K. 2013, \apj, 775, 73
\bibitem[Simon et al.(2015)]{sim15} Simon, J.B., Hughes, A.M., Flaherty, K.M., Bai, X.-N., \&\ Armitage, P.J. in prep
\bibitem[Semenov 
\& Wiebe(2011)]{sem11} Semenov, D., \& Wiebe, D.\ 2011, \apjs, 196, 25
\bibitem[Tilling et al.(2012)]{til12} Tilling, I., et al. 2012, \aap, 538, 20
\bibitem[Turner \& Drake(2009)]{tur09} Turner, N.~J., \& Drake, J.~F.\ 2009, \apj, 703, 2152
\bibitem[Turner et al.(2014)]{tur14} Turner, N.J., et al. 2014, PPVI
\bibitem[Visser et al.(2009)]{vis09} Visser, R., van Dishoeck, E.F., \&\ Black, J.H. 2009, \aap, 503, 323
\bibitem[Walsh et al.(2010)]{wal10} Walsh,C., Millar, T.J., \&\ Nomura, H. 2010, \apj, 722, 1607
 \bibitem[Walsh et al.(2014)]{wal14} Walsh, C., Juh{\'a}sz, 
A., Pinilla, P., et al.\ 2014, \apjl, 791, L6
\bibitem[Williams \&\ Best(2014)]{wil14} Williams, J.P., \&\ Best, W.M.J. 2014, \apj, 788, 59
\bibitem[Wilson(1999)]{wil99} Wilson, T.~L.\ 1999, Reports on 
Progress in Physics, 62, 143
\bibitem[Youdin 
\& Shu(2002)]{you02} Youdin, A.~N., \& Shu, F.~H.\ 2002, \apj, 580, 494
\end{thebibliography}
\end{document}